\documentclass[%
  aapm,
  mph,%
  amsmath,amssymb,
  reprint,%
]{revtex4-2}

\usepackage{xcolor}
\usepackage{graphicx}% Include figure files
\usepackage{dcolumn}% Align table columns on decimal point
\usepackage{bm}% bold math
\usepackage{hyperref}
\usepackage{url}
\usepackage{ulem}

\usepackage[mathlines]{lineno}% Enable numbering of text and display math

\newcommand{\rev}[1]{\textcolor{black}{#1}}

\begin{document}

\title{Unraveling the temporal dependence of ecological interaction measures}

\author{Javier Aguilar, Samir Suweis, Amos Maritan,  Sandro Azaele}
\affiliation{Dipartimento di Fisica e Astronomia Galileo Galilei, Universit{\`a} degli Studi di Padova, via Marzolo 8, 35131 Padova, Italy}

\date{\today}

\begin{abstract}
\rev{Identifying the network of species interactions is a fundamental step toward understanding ecosystem stability and biodiversity. However, the interpretability of empirical interaction measures remains a major challenge. Experimental estimates frequently exhibit puzzling temporal fluctuations, including sign shifts typically interpreted as transitions between competition and facilitation. Here, we analyze the temporal behavior of pairwise interaction measures to demonstrate that these fluctuations—and apparent shifts in ecological roles—can emerge intrinsically from standard population dynamics, without any underlying change in the actual ecological relationships. We show that inferred interactions are heavily distorted by experimental protocol choices, particularly the duration of observation and microbial growth constraints. By systematically evaluating interactions across timescales, we uncover a principled mechanism to mitigate these biases: short-term measurements reliably isolate direct, pairwise species couplings, whereas longer-term observations inevitably absorb indirect community feedbacks and systemic experimental constraints. By disentangling direct couplings from indirect network effects, our framework provides a robust, timescale-aware approach to interpreting empirical interaction matrices, offering critical quantitative guidance for experimental design and predictive ecosystem modeling.}
\end{abstract}

\maketitle

\section{Introduction}

%Understanding the behavior of a given species usually requires knowledge of its interplay with other members of its habitat.
The success of a species is rarely shaped in isolation; it emerges from a web of direct and indirect ecological interactions. Characterizing these interactions is challenging, as they involve complex relationships among multiple organisms~\cite{wootton_measurement_2005} that are shaped by a continuously changing environment~\cite{Yodzis1988} and eco-evolutionary feedbacks~\cite{Sanchez2013}. For example, the stress gradient hypothesis suggests that as environmental stress increases, facilitative interactions become more prominent while competitive interactions diminish~\cite{Bertness1994}. This implies that the environment not only determines the relative importance of species, but can also shift their roles between competitors and facilitators. Large-scale syntheses confirm broad empirical support for this prediction across taxa and ecological settings, with important implications for conservation under accelerating global change~\cite{Callaway2002,kefi2008,He2013,Adams2022,Diaz2024}. The difficulty of isolating clear relationships between community members has led to theories that either ignore species interactions altogether~\cite{Azaele2016,Grilli2020,SerGiacomi2018}, or model them as random variables~\cite{Allesina2015,kessler2015generalized,Akjouj2024,suweis2024generalized}. While such approaches have successfully explained some statistical patterns and broad ecological trends, constructing accurate, system-specific models—when feasible—requires robust empirical measurements of species interactions~\cite{Lawton1999,Volkov2009,fisher2014identifying,Foster2012}.

Correlations or co-occurrences across species are typically the first step in inferring interaction networks~\cite{Daleo2009,freilich2018species,Hervias2024}. Interaction measures are also commonly derived by fitting species abundance data with Lotka–Volterra or other population dynamics models, which are assumed to represent the underlying dynamics of the ecological communities~\cite{seifert_community_1976,faust2012microbial,Venturelli2018,Picot2023,Pasqualini2025}. Among the various model-free approaches to quantifying species interactions, estimates of the interaction matrix—defined as the effect of perturbations on a target species’ per-capita growth rate—are widely used~\cite{Bender1984,Laska1998,roxburgh_stability_2000,Attayde2001,Novak2016}. Empirical estimates of interaction matrices have quantitatively supported the stress-gradient hypothesis by detecting changes in the sign of interactions, which are associated with shifts between facilitative and competitive roles~\cite{Daleo2009,Armas2011,Piccardi2019,Martino2024}. However, a gap remains between the theoretical definition of interaction matrices—which assumes data of infinite resolution—and their empirical estimation, which is constrained by the finite resolution of measurements. Moreover, interaction measures do not inherently reveal whether species interactions are direct or mediated~\cite{wilson_application_1992,Menge1997}; such interpretations typically rely on contextual understanding of the biological system~\cite{kefi2015}. Overall, interaction measures may be biased not only by sampling limitations, environmental effects, and experimental design, but also by the theoretical framework used to interpret them. Such biases can obscure whether the measured interactions are direct or mediated, or whether they depend on the temporal or spatial scale of observation. This leads us to ask: how do empirical conditions affect interaction measures, and what kind of information do these measures truly convey? Is it possible to determine whether interactions are direct or mediated based on the measures alone?

\begin{figure*}
    \centering
    \includegraphics[width=1\linewidth]{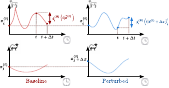}
    \caption{\rev{\textbf{Sketch of the experimental protocol to measure interactions with pulse experiments.} The axes represent the experiment duration ($t$), the population density of the perturbed species (bottom, depicted as a sheep, $x_j$ in the text), and the population density of the target species (top, depicted as a goat, $x_i$ in the text). The per-capita growth rates are measured in the curves of the top ($g_i^{(E)}$), while perturbations are introduces in the species of the bottom. Each element of the interaction matrix is calculated as the difference between the growth rates of the perturbed and baseline trajectories.}
}
\label{fig:sketch}
\end{figure*}

To address these questions, we will primarily study empirical interaction matrices extracted from synthetic data generated using paradigmatic ecological models. This approach allows us to clearly observe how interaction measures represent a perfectly known and controllable ecological process. Additionally, it enables the derivation of analytical results that provide general insights into the behavior of these interaction measures. We examine the impact of various empirical parameters, such as the resolution used to compute per-capita growth rate estimates, putting special focus on the experiment duration, i.e., the time elapsed from the ecosystem perturbation to the measurement of the per-capita growth rate. This parameter is particularly relevant in the analysis, as even if it is infinitesimal in the theoretical definition of the interaction matrix, it can be substantial in experimental setups~\cite{Menge1997,Attayde2001,Meacock2025}. Furthermore, it has been argued that long experimental durations reduce fluctuations in interaction estimates~\cite{Menge1997}. Through analytical calculations and simulations, we demonstrate that changes in the sign of the empirical interaction matrix can occur in purely competitive systems when the experiment duration is long enough. This is noteworthy because sign changes, traditionally linked to facilitation-competition trade-offs, may instead result from oscillations in species' relative abundances, even while their ecological roles remain fixed. 

Overall, we assess the critical role of experimental duration in characterizing ecological interactions. By temporally expanding the interaction matrix, we demonstrate that short-term measurements capture explicit couplings among ecological agents, whereas longer-term measurements increasingly reflect indirect feedback from the broader species community and environmental factors. Measuring interactions across different timescales therefore provides a more comprehensive view of system behavior: explicit species couplings can be inferred from short-term data, while indirect effects emerge over longer durations. In particular, we show that the nature of the experimental setup in microbial growth studies—such as chemostats, batch cultures, or biotic resource conditions—can significantly influence interaction estimates. Ultimately, we argue that interactions inferred at short durations, either through direct measurements or extrapolations from long-duration data, offer a systematic approach to distinguishing between direct and indirect interactions.

\section{The empirical interaction matrix}

\rev{In this section, we characterize the \rev{pulse} perturbation experiments and recall the standard definition of the empirical interaction matrix derived from them~\cite{Bender1984}.} Let us consider a general \rev{ecological community} consisting of $N$ species  with densities $x_i$, $i=1,\dots,N$. We denote by $x_i(t|\vec{x}^{(0)})$ the density of species $i$ at time $t$ given that the state of the ecological system at time $t=0$ was $\vec{x}^{(0)}$. Since we aim to analyze interactions in arbitrary ecological systems, the term ``species” may refer not only to organisms within a community but also to resources, either organic or inorganic. Likewise, ``densities” should be understood in a general sense, representing quantities such as biomass, abundance, or other relevant measures depending on the context. This information allows measuring the per-capita growth rate of species $i$ at time $t$, during the time interval $\Delta t$, as
\begin{equation}\label{eq:def_growth_rate_E}
    g^{(E)}_i\left[ \Delta t,t,\vec{x}^{(0)}\right] =\frac{x_i(t+\Delta t|\vec{x}^{(0)})-x_i(t|\vec{x}^{(0)})}{\Delta t \, x_i(t|\vec{x}^{(0)})}.
\end{equation}
In experimental settings, the per-capita growth rate can only be estimated at a finite set of time points $t_1, t_2, t_3, \dots$ where species densities are measured. Consequently, in real applications, the parameter $\Delta t$ in \eqref{eq:def_growth_rate_E} is the temporal resolution of the time series, $\Delta t_j = t_{j+1} - t_j$.

\rev{Pulse} experiments are based on having access to (at least) two time series of the same ecological system with different initial conditions. Specifically, the empirical interaction matrix is estimated by comparing pairs of time series, where the difference in the per-capita growth rate is evaluated between a baseline (unperturbed) evolution and a perturbed evolution. In the perturbed evolution, the initial density of species $j$ is altered by an amount $\Delta x$, allowing us to assess the impact of this change on the system's dynamics,
\begin{align}\label{eq:def_interact_E}
    \mathcal{M}_{j\to i}&\equiv\mathcal{M}_{i,j} (\vec{x}^{(0)},t,\Delta x,\Delta t)\nonumber \\
    &=g^{(E)}_i\left[\Delta t,t,\vec{x}^{(0)}+\Delta x \,\hat{j}\right]-g^{(E)}_i\left[\Delta t,t,\vec{x}^{(0)}\right],
\end{align}
where $\hat{j}$ is the unit vector in the $j$-th direction.  \rev{In Fig.~\ref{fig:sketch} we show a graphical representation of the baseline and perturbed experiments, together with experimental parameters.} 

\rev{The interactions measured via Eq.~\eqref{eq:def_interact_E} echo the central idea of linear response theory, namely that small perturbations can be used to extract information about the structure and stability of physical~\cite{marconi_2008} and ecological~\cite{goyal_universal_2025} systems. However, Eq.~\eqref{eq:def_interact_E} reflects how perturbations would be implemented in real experiments, where practical constraints such as finite sampling times and perturbation sizes become relevant. In this sense, the matrix in Eq.~\eqref{eq:def_interact_E} is the \emph{empirical interaction matrix}, which is intended to capture ecological interactions (e.g., facilitation, competition) and may differ from the coupling parameters appearing in ecological models.  }

\rev{The matrix in Eq.~\eqref{eq:def_interact_E} provides direct insight into the relationships between species:} the sign of $\mathcal{M}_{j \to i}$ indicates whether changes in species $j$ promote or inhibit the growth of species $i$, while its magnitude reflects the strength of this influence. However, $\mathcal{M}_{j \to i}$ also depends on experimental parameters that are not intrinsic to ecological dynamics. In particular, it is affected by the perturbation size $\Delta x$, the total experiment duration $t$, and the sampling interval $\Delta t$. When $ t$ is large, we will show that $\mathcal{M}_{j \to i}$ captures more of the system-wide feedback, incorporating indirect effects. In contrast, for small $t$, it predominantly reflects the direct impact of species $j$ on species $i$, providing a clearer measure of their direct interaction, with minimal contributions from the rest of the system. 

\rev{Although measures of interaction matrices based on pulse experiments are theoretically well founded and potentially highly informative, their empirical implementation remains challenging. In particular, perturbing natural ecosystems in a controlled and targeted way is rarely feasible, and ecological data are often sparsely sampled in time. Moreover, experimental perturbations that are treated as infinitesimal in theory are necessarily finite in practice~\cite{Bender1984, Attayde2001}, creating a gap between theoretical studies of interaction matrices~\cite{Bender1984,Attayde2001,Novak2016} and their empirical estimation. These constraints are further compounded by the difficulty of applying unidirectional perturbations and obtaining high–temporal-resolution measurements of system responses. This work therefore focuses on understanding how experimental parameters influence the information that can be extracted from $\mathcal{M}_{j \to i}$.}
\begin{figure*}
    \centering
    \includegraphics[width=1\linewidth]{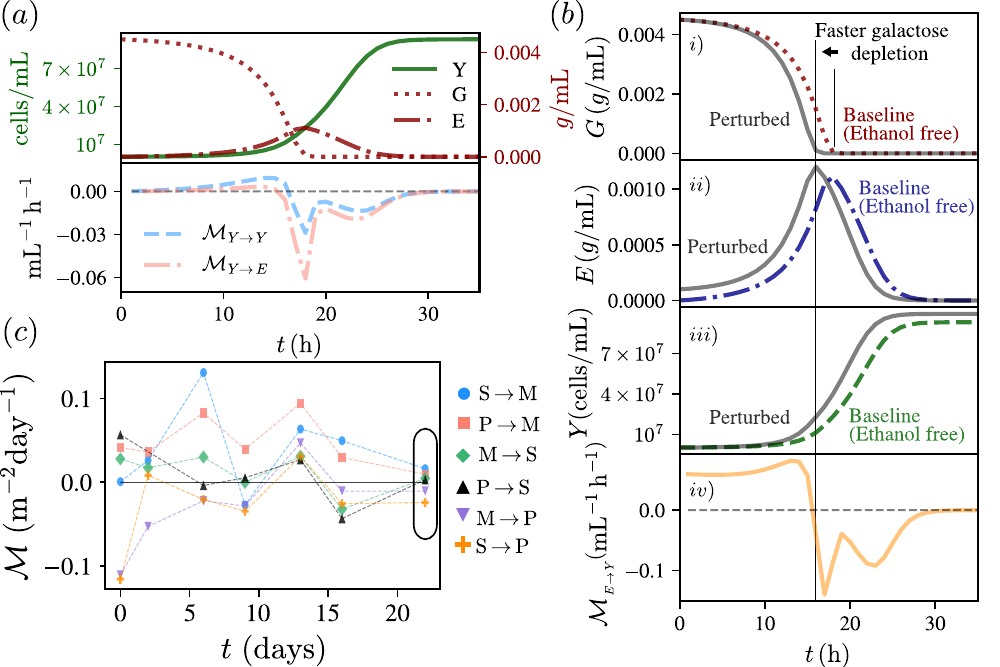}
    \caption{\textbf{Temporal variability of interaction measures.} (a, top): Temporal evolution of yeast (Y), galactose (G), and ethanol (E) densities in the baseline trajectories. (a, bottom): Temporal evolution of interactions induced by yeast perturbations. \rev{(b-i): Galactose dynamics in the baseline trajectory—initially ethanol-free—and in the perturbed trajectory, which includes ethanol from the start (dotted and solid lines, respectively). (b-ii): Similar to b-i but showing ethanol evolution. (b-iii): Yeast population dynamics with and without the initial presence of ethanol. (b-iv): Evolution of the interaction measure $\mathcal{M}_{E\to Y}$.} (c): Temporal evolution of inferred interactions among grasshoppers—\emph{M. femur-rubrum} (M), \emph{S. collare} (S), and \emph{P. nebrascensis} (P). Details for parameters and model used to generate a,b in Sec.~\ref{sec:parameters}.
}
    \label{fig:data}
\end{figure*}

\section{Results}

\subsection{Temporal fluctuations in empirical interaction measures: evidence from microbial and consumer systems}

Here, we apply the interaction measure presented in Eq.~\eqref{eq:def_interact_E} to two distinct \rev{consumer-resource systems} to illustrate the significance of temporal fluctuations in interactions. The first system involves the temporal dynamics of a yeast culture (S. cerevisiae) grown in a medium initially containing galactose as the sole carbon source. As the yeast consumes galactose, it produces ethanol as a byproduct, which can subsequently be used as a secondary resource~\cite{Pacciani-Mori2020}. This system thus represents a simple \rev{consumer-resource system} composed of a single species interacting with two resources. We employ a consumer-resource model, fitted to experimental data from ref.\cite{Pacciani-Mori2020}, to generate time series from which the interaction matrix is computed using Eqs.\eqref{eq:def_growth_rate_E} and~\eqref{eq:def_interact_E}. Since we can generate time series with arbitrary resolution, we focus on the effect of the experiment duration ($t$) in the limit of small perturbation and sampling time ($\Delta x,\Delta t\to 0$). Figs.~\ref{fig:data}a,b illustrate the evolution of interactions, which includes sign changes over time. Fluctuations in Fig.~\ref{fig:data}a reflect changes in the relevance of metabolic pathways. In the early stages, ethanol production exceeds consumption, resulting in a net positive interaction between ethanol and yeast. Similarly, during this initial phase, the benefits of ethanol production outweigh any competitive pressures, leading to initially positive yeast self-interactions. However, this situation changes dramatically as galactose becomes depleted. Once the system runs out of galactose, ethanol consumption surpasses production, and further increases in yeast concentration begin to inhibit both yeast growth and ethanol levels. In Fig.\ref{fig:data}b, we examine how ethanol perturbations affect yeast growth rates ($\mathcal{M}_{E\to Y}$). This setup compares yeast growing in a  dilution containing only galactose --baseline experiment-- with another evolution where a small amount of ethanol is also present --perturbed experiment--. Since resource addition typically promotes consumer growth, $\mathcal{M}_{E\to Y}$ is initially positive, and yeast densities are higher in solutions containing both ethanol and galactose than in those with only galactose (see \rev{Fig.\ref{fig:data}b-iii)}. However, the perturbed condition depletes galactose more quickly, entering faster the phase where yeast growth relies solely on ethanol (see Fig.\ref{fig:data}b-i), which supports a lower growth rate. This dynamic explains the sign shift in \rev{Fig.~\ref{fig:data}b-iv} and illustrates how resource addition can eventually lead to negative interactions with consumers.

The second dataset consists of empirical measurements of the population dynamics of three grasshopper species (\emph{Melanoplus femur-rubrum}, \emph{Spharagemon collare}, and \emph{Phoetaliotes nebrascensis}) maintained in captivity. The data originate from experiments reported in ref.~\cite{Ritchie1993}, in which enclosures were placed in areas of natural grassland. Each enclosure contained native vegetation (serving as the grasshoppers’ shared resource) and was initialized with different population densities. \rev{Grasshoppers are expected to compete for common resources. The varying initial conditions are treated as perturbations that  allow us to infer interspecific interactions (see Section ~\ref{sec:grasshoppers} for details and confidence intervals).} In Fig.~\ref{fig:data}c, we show that the inferred interactions oscillate over time, with many of them changing sign during the course of the experiment. Such sign changes were not reported in the original study, which only analyzed interactions at the final time point of the experiment ($t = 28$ days). That analysis was interpreted through a biological lens: dietary niche overlap among grasshoppers and possible indirect positive interactions with plants —mediated by soil enrichment— were proposed to explain the observed interaction patterns. It is noteworthy that this interpretation could be significantly affected when considering interactions at intermediate experiment durations.

Through these examples, we have examined the emergence of fluctuating interaction measures in both microscopical and macroscopical \rev{consumer-resource systems}. Understanding the meaning of interaction sign changes in the yeast–galactose–ethanol system  required a great level of knowledge of organism dynamics, which was not possible to apply in the grasshopper system, which is more difficult to interpret as it is affected by noise.

These results illustrate that variability in interaction measures can arise from multiple sources in both microscopic and macroscopic ecosystems. In Fig.\ref{fig:data}a, we observe that the shifts between facilitation and competition correspond to changes in the sign of interactions, consistent with previous findings\cite{Piccardi2019,Picot2023,Meacock2025}. In contrast, fluctuations observed in the grasshopper system (Fig.\ref{fig:data}c) are harder to interpret. These may stem from the intrinsic randomness of the real-world data and the complex dynamics of the system as a whole. More striking is the behavior seen in Fig.\ref{fig:data}b: although resource addition typically promotes consumer growth —suggesting consistently positive interactions— we observe sign shifts that defy this expectation. In the following section, we demonstrate that such sign changes are not anomalies but a general feature, even in deterministic (error-free) models of purely {\it competitive} ecological systems. 

\subsection{A general framework for computing interaction measures from synthetic data}\label{sec:computing_interactions}

In order to ensure total controllability of the ecosystem's dynamics, we will study interaction measures in synthetic time series generated with paradigmatic ecological models. In particular, we focus on deterministic models defined through the instantaneous per-capita growth rate
\begin{equation}\label{eq:def_model} 
    \dot{x}_i = x_i\,g_i\left(\vec{x}\right). 
\end{equation}
By integrating Eq.~\eqref{eq:def_model}, we can generate time series starting from any initial condition $\vec{x}^{(0)}$ and with arbitrary sampling times. Evidently, the empirical estimations of the per-capita growth rate (given by Eq.~\eqref{eq:def_growth_rate_E}) converge to the instantaneous growth rate that defines the model in the limit of small $\Delta t$,
\begin{equation}\label{eq:limit_empirical_growth_rate}
    \lim_{\Delta t \to 0} g^{(E)}_i \left( \Delta t,t,\vec{x}^{(0)}\right) = g_i\left( \vec x(t|\vec{x}^{(0)})\right).
\end{equation}
We note that Eqs.~\eqref{eq:def_growth_rate_E} and~\eqref{eq:def_interact_E} make no model assumptions, since they only depend on raw data. The model in Eq.~\eqref{eq:def_model} just provides the time series where Eq.\eqref{eq:def_interact_E} is evaluated.  This approach also allows to compute the value of $\mathcal{M}_{i,j}$ analytically to gain general insights into the effects of $\Delta t$, $\Delta x$, and $t$. In real experiments, much care is taken to keep the sampling time and perturbation magnitude small relative to the scales of density fluctuations~\cite{Novak2016}; however, the duration of the experiment can be relatively long. Long durations of experiments have also been claimed to decrease fluctuations of interaction estimates~\cite{Attayde2001}. Therefore, we perform a Taylor expansion to Eq.\eqref{eq:def_interact_E} to obtain interactions using small $\Delta x$ and $\Delta t$ while maintaining arbitrary $t$ (see details in Appendix~\ref{AP_sec:proof_formula_interactions}). In doing so, we find a scalar product form of a vector only dependent on species $i$ and another one only dependent on species $j$,
\begin{align}\label{eq:interactions_computed_with_the_model}
    \mathcal{M}_{i,j} = \vec{G}_i\cdot \vec{E}_j \, \Delta x +\mathcal{O}(\Delta t)+\mathcal{O}(\Delta x^2),
\end{align}
where we defined the gradient vector
\begin{equation}
    \vec{G}_i = \nabla\cdot g_i(\vec{x})\Bigg|_{\vec{x}=\vec{x}(t|\vec{x}^{(0)})}  ,
\end{equation}
and the evolution vector
\begin{equation}
    \vec{E}_j = \partial_{x^{(0)}_j} \vec x(t|\vec{x}^{(0)}),
\end{equation}

Therefore, species $j$ facilitates (respectively hinders) the growth of species $i$ when the evolution vector $\vec{E}$ aligns with (respectively opposes) the direction of maximum variation in the growth rate of species $i$ at time $t$. 

\rev{\rev{Eq.~\eqref{eq:interactions_computed_with_the_model} allows us to relate the interaction matrix $\mathcal{M}$ derived from pulse perturbation experiments to other commonly used measures of species interactions. In particular, in the limit of short experiment times the interaction matrix reduces to the coupling constants of Lotka--Volterra models (see Section~\ref{sec:model_inference}), and therefore coincides with the interaction coefficients obtained through regression methods applied to such models (e.g. Ref.~\cite{seifert_community_1976}). This shows that interaction measures based on pulse perturbations recover the standard Lotka--Volterra interaction coefficients in the appropriate limit. Moreover, Eq.~\eqref{eq:interactions_computed_with_the_model} generalizes similar expressions previously derived for environment--organism models in Ref.~\cite{Meacock2025} (see Appendix~\ref{AP_sec:OE_model}).}}

Shifts in self-interactions for single-species models are straightforward to analyze, since the interaction sign coincides with the sign of the gradient vector, which reduces to a scalar in this setting (see proof in Appendix~\ref{AP:One_dim_int}):
\begin{equation} 
\text{sign} \left( \mathcal{M}\right) = \text{sign} \left( G \right). 
\end{equation}
However, in multi-species models, the interaction sign depends on the evolution of both the gradient $\vec{G}_i$, which is typically computable, and the vector $\vec{E}_j$, which is more difficult to evaluate analytically. This difficulty arises because computing derivatives with respect to $x_k^{(0)}$ requires explicit knowledge of how the system's trajectories depend on initial conditions. In practice, this means solving the full dynamics, which is generally infeasible for most systems. Therefore, we will approximate the vector $\vec{E}_j$ in specific time regimes to gain analytical insight.

\subsection{Interaction measures exhibit intrinsic temporal fluctuations and sign shifts even in pure competitive systems}

\begin{figure}
    \centering
    \includegraphics[width=1.05\linewidth]{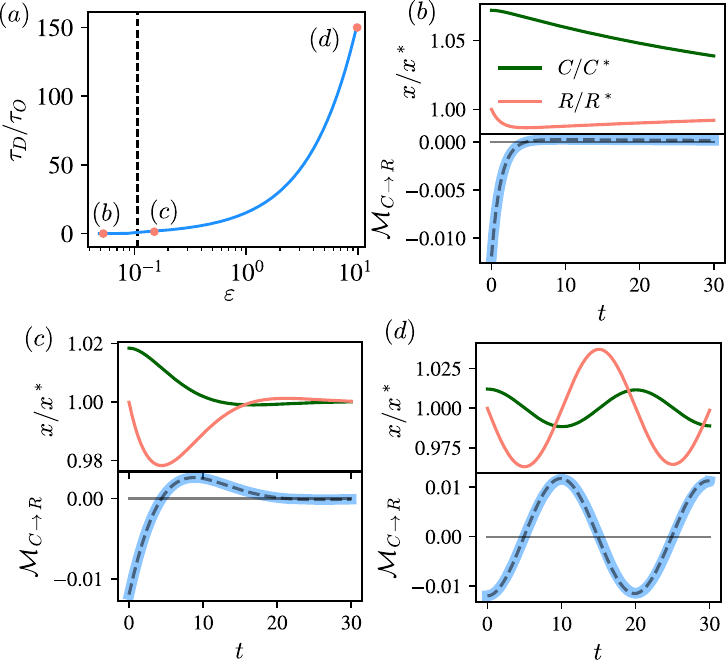}
    \caption{ \rev{\textbf{Interaction sign changes in purely competitive systems: one consumer-one resource model.}} \rev{In (a), ratio between damping and oscillating time-scales in the one consumer-one resource model with biotic resource [Eq.\eqref{eq:1C_1R_CR_R_eq}] for different values of the consumer uptake efficiency. Vertical dashed line signals the threshold $\varepsilon_\text{TH}$ below which there are no oscillations in the relaxation toward equilibrium. Dots correspond to the data used in (b), (c), and (d). In (b), (c), and (d) we show instances of perturbed trajectories (top) and the associated measure of interactions (bottom), with  continuous lines corresponding to interactions computed from raw data, while dashed line corresponds to analytical results (Eqs.~\eqref{eq:interactions_computed_with_CR_model_linearized} and~\eqref{eq:interactions_GBM_model}).  In (b) monotonic relaxation corresponds to interaction measures that do not change sign. In (c) small fluctuations rapidly damped exhibit changes of sign in the interactions. In (d), relaxation dominated by oscillations corresponds to wild oscillations in the interaction measures. $R^*$ and $C^*$ notes, respectively, to the densities of consumers and resources at equilibrium.  See parameter values in Sec.~\ref{sec:parameters}.}}
    \label{fig:interactions_1C1R}
\end{figure}

\begin{figure*}
    \centering
    \includegraphics[width=1.05\linewidth]{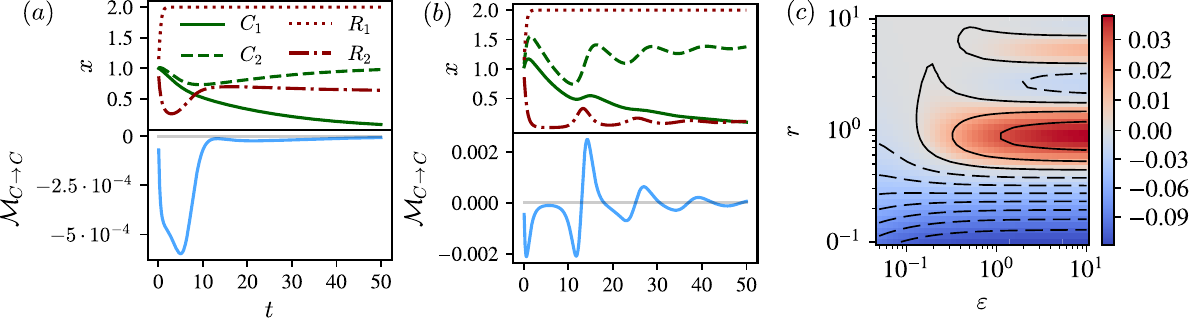}
    \caption{ \rev{\textbf{Interaction sign changes in purely competitive systems: multiple consumers and resources.}} \rev{Multi-species data can either show consistent interaction measures that do not change sign (a) or erratic interaction sign switches (b). In (c), integral of interactions in a time window according to Eq.~\eqref{eq:I_M_RC} for different values of the consumer and resource uptake efficiency ($\varepsilon$ and $r$ respectively) also exhibit different signs depending on the parameters values. Synthetic datasets and parameter choices available in Sec.~\ref{sec:parameters}.}}
    \label{fig:competitive_systems}
\end{figure*}

We focus on synthetic datasets generated with multi-species consumer–resource models without facilitation, meaning that simulated species only compete for limited resources, which themselves follow their own dynamics. Specifically, we consider a classical consumer-resource model with a general renewal function for the resources~\cite{MacArthur1970,Chesson1990,Ritchie1993,Palamara2021},
\begin{equation}\label{eq:model_CR_C}
    \dot{C}_i = C_i \left( \varepsilon_i\sum_{\ell}^{n_R} \alpha_{i,\ell} R_\ell - d_i\right),
\end{equation}
and
\begin{equation}\label{eq:model_CR_R}
    \dot{R}_\ell = h_\ell(\vec{R})-R_\ell\sum_{j}^{n_C} C_j \alpha_{j,\ell} .
\end{equation}
Where $n_R$ and $n_C$ are the number of resources and species respectively, $\alpha_{i,\ell}$ is the rate of consumption of resource $\ell$ by species $i$, $\varepsilon_i$ is the consumer's uptake efficiency, and $d_i$ is the death rate. The renewal function $h_\ell$ encodes the rate at which resource $\ell$ enters the system. All parameters and the renewal function are considered to be non-negative. \rev{This modeling framework is particularly well suited for microbial communities, where pulse perturbations are easier to implement experimentally and spatial structure often plays a less important role compared to macroscopic ecosystems.}

\rev{Using dynamical systems theory~\cite{Strogatz2018,Zoltan}, we show that empirical interactions measured between consumers and resources in these systems may differ from the coupling constants $\alpha$, which encode the direct coupling between them. More importantly, we demonstrate that changes in the sign of interactions among consumers can occur in systems with an arbitrary number of consumers and resources (see details in Sec.~\ref{sec:measuring_interactions_LRT}). Consequently, interaction measures may switch sign even in purely competitive ecological systems.}

To illustrate this result, we analyze interactions in a simpler setting consisting of
one consumer and one resource characterized by
\begin{equation}\label{eq:1C_1R_CR_R_eq}
   \frac{\dot{C}}{C} = \varepsilon\,  \alpha\,R-   d,\quad\text{and} \quad  \frac{\dot{R}}{R} = r  \left(1-\frac{R}{K}\right)-   \alpha\,  C.
\end{equation}
For \( r > 0 \), this model describes a biotic resource undergoing logistic growth. Thus, this version of the consumer–resource model can also be interpreted as a predator–prey system, where the prey (resource) is depleted by the predator (consumer). \rev{The model admits a single coexisting stationary state. We focus on the regime in which this fixed point is stable. In this regime, small perturbations around the stationary state decay over time and the system returns to equilibrium.} The timescales governing this relaxation are determined by two key quantities: the characteristic damping timescale $\tau_D = \left| \text{Re}(\lambda_+) \right|^{-1}$, and the oscillation timescale, 
$\tau_O = \left| \text{Im}(\lambda_+) \right|^{-1}$, 
where \( \lambda_+ \) is the eigenvalue of the Jacobian matrix with the largest real part. \rev{It is well-known that this model displays a threshold tuning the appearance of oscillations~\cite{rosenzweig_paradox_1971,may_limit_1972}. Specifically,} when
\begin{equation}\label{eq:threshold_1C_1R}
\varepsilon < \varepsilon_{\text{TH}} = \frac{\sqrt{dr + d^2} + d}{2\alpha K},
\end{equation}
the system relaxes without oscillations \((\tau_O^{-1} = 0)\), and the dynamics are dominated by exponential damping (see Fig.~\ref{fig:competitive_systems}a). \rev{See details of this analysis in Appendix~\ref{AP_sec:1C1R}.}

\rev{Intuitively, the interaction $\mathcal{M}_{C\to R}$, i.e., how perturbations in $C$ affect the growth rate of $R$, should be negative, reflecting that the addition of a consumer hinders resource growth. However, by varying the consumer's uptake efficiency ($\varepsilon$), we observe a range of interaction scenarios—from predominantly negative (Fig.~\ref{fig:interactions_1C1R}b), to weakly oscillatory (Fig.~\ref{fig:interactions_1C1R}c), to strongly oscillatory (Fig.~\ref{fig:interactions_1C1R}d). This variability arises because $\varepsilon$ not only controls the onset of oscillations through Eq.~\eqref{eq:threshold_1C_1R}, but also determines the relative importance of damping and oscillatory time scales, as shown in Appendix~\ref{AP_sec:1C1R}.When population oscillations are present, fluctuations in interaction measures reflect the inherently oscillatory nature of the population dynamics. Therefore, linear analysis at the population level, which has been used to study ecological responses to perturbations~\cite{goyal_universal_2025}, can also be used to quantitatively explain the observed variability of interaction measures.}

Overall, this simple example demonstrates that interaction measures form a dynamical system in their own right, with temporal dynamics that can be analyzed similarly to species dynamics in traditional population models. The relevance of fluctuations in these measures depends on the biological characteristics of the system — or, in the case of synthetic data, on the underlying model parameters. In multi species consumer–resource \rev{systems}, when there is a clear separation of timescales between consumers and resources~\cite{Chesson1990}, fluctuations driven by population dynamics become negligible (\rev{see Fig.~\ref{fig:competitive_systems}a}). In contrast, significant fluctuations\rev{, both in species densities and interaction measures,} are expected when the temporal scales of consumers and resources are comparable (\rev{see Fig.~\ref{fig:competitive_systems}b}). In \rev{Fig.~\ref{fig:competitive_systems}c}, we show that the temporal integral of interactions over a finite time window $[0,t]$,
\begin{equation}\label{eq:I_M_RC}
    I_{R\to C} (\vec{x},t,\Delta t,\Delta x) = \frac{1}{t}\int_0^t \mathcal M_{R\to C} (\vec{x},s,\Delta t,\Delta x)\,ds,
\end{equation}
exhibits different dominant signs depending on the system's parameters. Therefore, sign switches are not ‘averaged out’ through temporal integration; depending on these parameters, interactions can be predominantly negative or positive within a fixed time window.

\subsection{Short-time expansion reveals the distinction between direct and indirect interactions}\label{sec:interactions_close_to_perturbation}

\begin{figure*}
    \centering
    \includegraphics[width=1\linewidth]{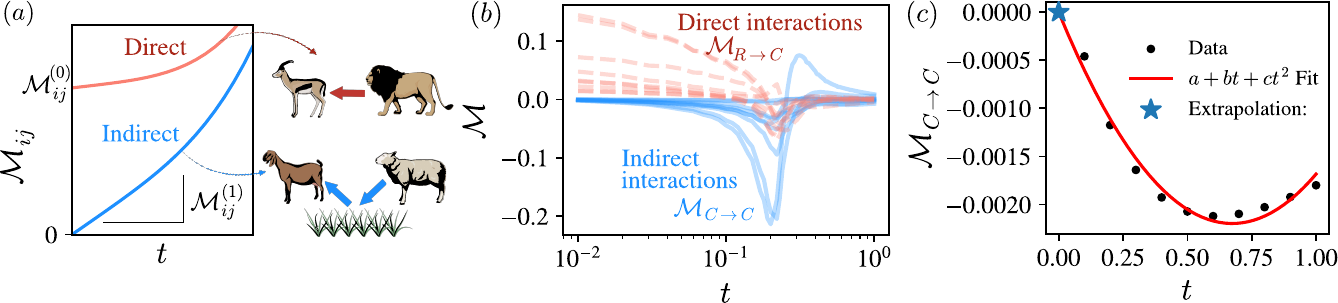}
    \caption{\rev{\textbf{Temporal dependence disentangles direct and indirect interactions .}} In a, sketch illustrating the hallmarks of direct and indirect interactions. In the limit \( t \to 0 \), direct interactions approach a finite value (\( \mathcal{M}^{(0)}_{i,j} \)), while indirect interactions vanish proportionally to \( \mathcal{M}^{(1)}_{i,j}\,t \). In b, interaction measures for short experiment durations in a multi-species consumer-resource model. The interactions \( \mathcal{M}_{C \to C} \) tend to zero, while \( \mathcal{M}_{R \to C} \) converge to a positive value as \( t \to 0 \). In c, interaction measures between consumers at various discrete experiment durations (dots); data is fitted with a parabolic function (solid line), and extrapolation to \( t = 0 \) correctly recovers the interaction limit (\( \mathcal{M}^{(0)}_{C \to C} = 0 \)).   Synthetic datasets and parameter choices available in Sec.~\ref{sec:parameters}.}
    \label{fig:scaling}
\end{figure*}

In the previous section, we demonstrated that sign changes in interaction measures can arise from intrinsic oscillations during relaxation processes. This presents challenges for interpreting the outcomes of \rev{pulse} perturbative experiments, as identical species and environmental conditions can lead to significantly different results depending on the experiment's duration. Consequently, a natural question arises: what experiment duration should practitioners use to ensure that interaction measures are informative? In the following, we show that by performing interaction measurements at multiple experiment durations, one can obtain a more comprehensive understanding of system dynamics, capturing not only species-specific behavior but also their interactions with the environment and the experimental setup. To do so, we turn our attention to interactions occurring over short experiment durations. This analysis is carried out through the expansion of Eq.~\eqref{eq:interactions_computed_with_the_model} as a power series in the experiment duration, $t$, and in the limit of vanishing $\Delta t$,
\begin{equation}\label{eq:expansions_interactions}
    \mathcal{M}_{i,j} (\vec{x},t) = \mathcal{M}_{i,j}^{(0)} (\vec{x})+ \mathcal{M}_{i,j}^{(1)} (\vec{x})\, t +\mathcal{M}_{i,j}^{(2)} (\vec{x})\, t^2+\dots
\end{equation}
The zeroth-order in the expansion simply becomes the derivative of the per-capita growth rate evaluated at the perturbation point (see Appendix~\ref{AP_sec:Expansion_experiment_duration} for details):
\begin{equation}\label{eq:zeroth-order_approx_M}
    \mathcal{M}_{i,j}^{(0)} (\vec{x}) = \frac{\partial}{\partial x_j} g_i(\vec{x}) \, \Delta x.
\end{equation}
Therefore, the zeroth-order term in the expansion depends solely on the gradient vector $\vec{G}_i$ and captures the explicit couplings between species $i$ and $j$ present in the model used to generate the synthetic data. As such, this term represents the direct interactions between species. In contrast, the higher order terms in time in Eq.~\eqref{eq:expansions_interactions} depend on both $\vec{G}_i$ and $\vec{E}_i$, and they encode the effects of indirect interactions among species, environmental dynamics, and experimental protocols. Specifically, if there is no explicit coupling between species $i$ and $j$, i.e., if $\frac{\partial}{\partial x_j} g_i(\vec{x}) = 0,$ then the leading term in the expansion becomes (see Appendix~\ref{AP_sec:Expansion_experiment_duration}):
\begin{equation}\label{eq:first-order_approx_M}
    \mathcal{M}_{i,j}^{(1)}(\vec{x}) = \nabla\left(\partial_{x_j}g_i(\vec{x})\right)\cdot\dot{\vec{x}} + \nabla g_i(\vec{x})\cdot\partial_{x_j}\dot{\vec{x}}.
\end{equation}
For a broad class of systems, Eq.~\eqref{eq:first-order_approx_M} simplifies, and $\mathcal{M}_{i,j}^{(1)}(\vec{x})$ coincides with the instantaneous interaction matrix defined in Ref.~\cite{Meacock2025}, which characterizes indirect interactions among species mediated by the environment (see Appendix~\ref{AP_sec:OE_model}).

The above analysis provides a method to determine whether an interaction measure reflects direct or indirect interactions between species. The key idea is to exploit the fact that these two types of interactions behave differently in the limit $t \to 0$. While indirect interactions should tend to zero near $t \approx 0$, direct interactions should intercept the y-axis at a finite value corresponding to $\mathcal{M}_{ij}^{(0)}$ (see sketch in Fig.~\ref{fig:scaling}a). In Fig.~\ref{fig:scaling}b, we show the different behavior of direct and indirect interactions using a multi-species consumer resource model. The interactions $\mathcal{M}_{R \to C}$—which quantify how resource perturbations affect consumer growth—tend toward a positive value as $t \to 0$, reflecting the explicit coupling between consumers and resources. In contrast, the interactions $\mathcal{M}_{C \to C}$ tend towards zero, indicating that the consumer-consumer interactions are indirect and mediated by resource availability. In Fig.~\ref{fig:scaling}c, we show the case where $\mathcal{M}_{C \to C}$ interactions are only available at finite experimental times. Although measures with finite $t$ are always finite and negative, reflecting mediated competition, the uncoupled nature of consumers is revealed by the limit $t\to 0$, which is accessed through extrapolation. \rev{In practice, we use a parabolic fit, which allows us to directly match the coefficients of the fit with the terms of the series expansion of the interaction matrix (Eq.~\eqref{eq:expansions_interactions}). Other extrapolation schemes, such as higher-order polynomial fits or spline interpolation, could also be used to estimate the $t\to0$ limit.}

\subsection{Experimental protocols shape interaction measures through resource dynamics}\label{sec:Effect_of_experiment}

\begin{figure}
    \centering
    \includegraphics[width=0.8\linewidth]{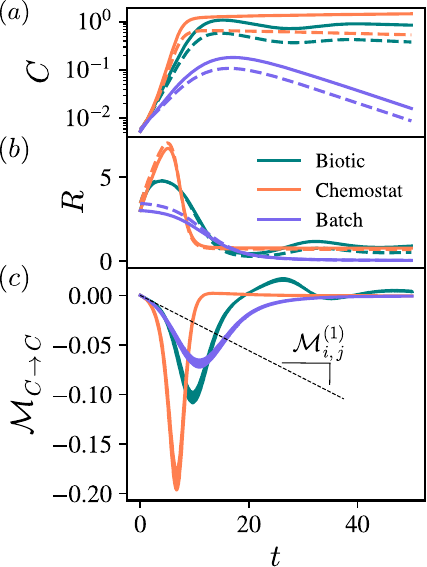}
    \caption{\rev{\textbf{Renewal type impacts measures of interactions.} In (a), baseline (continuous) and perturbed (dashed) trajectories of consumers in a two-consumer, two-resource model, with different colors representing resource renewal types (biotic, chemostat, batch), while coupling constants are identical across all types.  In (b), similar to (a), but showing trajectories of resource dynamics. In (c), interaction measures across renewal types corresponding to the trajectories in (a).}}
    \label{fig:experiment_influence}
\end{figure}
In experimental realizations of consumer–resource communities—both those involving macroscopic organisms and microbial communities—the experimental protocol is often largely defined by the nature of resource input into the system. One approach is to sample species’ growth rates under natural resource conditions (see e.g.~\cite{Menge1997,Paine1974}). To achieve more controlled conditions, experimentalists can instead isolate communities from their natural environments and (i) study population decay when only an initial quantity of resources is provided without external input~\cite{Ritchie1993,Foster2012,Goldford2018,Neythen2020}, or (ii) provide a continuous input of selected resources to investigate stationary properties~\cite{Vetsigian2011,Venturelli2018,Picot2023}. Since we have shown that population dynamics influence the outcomes of interaction measures, we expect that the choice of experimental protocol may significantly affect these measures. In the consumer-resource model defined in Eqs~\eqref{eq:model_CR_C} and~\eqref{eq:model_CR_R}, different experimental protocols can be analyzed through the choice of the resource renewal function $h_\ell$. In particular, we focus on three types of resource renewal that can be found in experiments of microbial communities:
\begin{itemize}
    \item \textbf{Batch culture:} There is no renewal of resources. An initial quantity of resources is present and decays due to consumption. Thus, \makebox{$h_\ell(\vec{R}) = 0$}.
    \item \textbf{Chemostat:} Abiotic resources are introduced at a constant rate, establishing a steady state. This corresponds to a constant renewal function, \makebox{$h_\ell(\vec{R}) = \Gamma_\ell$}.
    \item \textbf{Biotic resource:} Resources follow their own natural dynamics and coevolve with the consumers. We model this with logistic growth, \makebox{$h_\ell(\vec{R}) = r_\ell R_\ell(1 - R_\ell / K_\ell)$}.
\end{itemize}
By computing the explicit parameter dependence of the expansion in Eq.~\eqref{eq:expansions_interactions}, we find that the first two terms behave identically across all experimental protocols, as they do not depend on the resource renewal functions or on the consumers' death rates (see Appendix~\ref{AP_sec:CR_model}):
\begin{equation}\label{eq:expansion_CR}
    \begin{cases}
        \mathcal{M}_{C_j \to C_i}^{(0)} = 0,\\
        \mathcal{M}_{C_j \to C_i}^{(1)} = -\varepsilon_i \sum_{\ell=1}^{n_R} R_\ell \,\alpha_{i,\ell} \,\alpha_{j,\ell}.
    \end{cases}
\end{equation}
In Eq.~\eqref{eq:expansion_CR}, $\mathcal{M}^{(0)}$ reflects the absence of direct interactions among consumers, while $\mathcal{M}^{(1)}$ captures the expected negative indirect interactions mediated by resource competition. Notably, $\mathcal{M}^{(1)}$ generalizes the interaction structure found in the Lotka–Volterra approximation of the consumer–resource model with fast resource dynamics\cite{Chesson1990}.  In Appendix~\ref{AP_sec:CR_model} we show that the next term in the expansion, $\mathcal{M}_{C_j \to C_i}^{(2)}$, already depends on the renewal function and can be either positive and negative. \rev{In Figs.~\ref{fig:experiment_influence}} we show trajectories and interactions in a two-consumer, two-resource model with different resource renewal functions but identical consumer-resource coupling constants, thus representing constant consumer-resource relationships under different experimental conditions. Interaction measures in all experimental settings behave similarly at very short times, when consumer-resource coupling dominates, in line with Eq.~\eqref{eq:expansion_CR}. However, as the experiment progresses, the impact of the experimental protocol on interaction measures grows. Additionally, we show that changes of sign in the interaction could critically depend on the renewal function, suggesting that sign changes could be artifacts induced by the experimental setting.

\subsection{Model inference from empirical interaction measures}\label{sec:model_inference}
\begin{figure}
    \centering
    \includegraphics[width=0.9\linewidth]{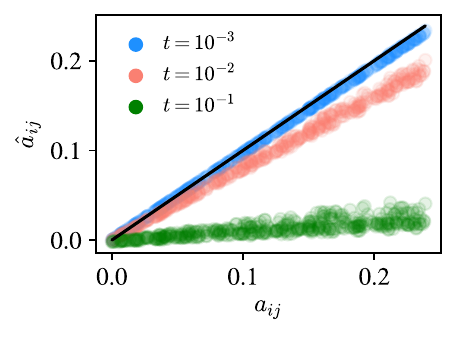}
    \caption{\rev{\textbf{Parametric inference from empirical interaction measures .} Inference of coupling constants ($\alpha$'s in Eq.~\eqref{eq:model_CR_C}) in a 20-consumer, 20-resource model using interaction measures with three different experiment durations. The x-axis shows true parameter values, the y-axis shows inferred values, and the solid line indicates perfect inference.  Parameter choices for synthetic dataset available in Sec.~\ref{sec:parameters}.}  }
    \label{fig:inference}
\end{figure}
So far, we have demonstrated how empirical estimates of the interaction matrix behave on synthetic data generated by models. However, we could also ask which types of model can be inferred from measurements of the interaction matrix. Using the normalized version of the interaction matrix~\cite{Novak2016},
\begin{equation}
    \hat{a}_{ij}(\vec{x}) = \frac{1}{\Delta x} \mathcal{M}_{i,j} (\vec{x},t,\Delta x,\Delta t),
\end{equation}
then the zeroth order model inferred from interactions is
\begin{equation}\label{eq:model_from_interactions}
    \dot{x}_i = x_i \sum_j \int^{x_j}_0 \hat{a}_{ij} \left(x_1,\dots,x_{j-1},s,x_{j+1},\dots\right)  ds + x_i c_i.
\end{equation}
The derivation of Eq.\eqref{eq:model_from_interactions}, detailed in Appendix~\ref{AP_sec:inferring_model_from_interactions}, relies on the fact that in the limit $t\to0$, the relationship between interaction measures and per-capita growth rates becomes trivial, \rev{and model inference becomes similar to Jacobian reconstruction approaches~\cite{barter_2021}.} This inference method cannot capture one-body processes (such as death, asexual reproduction, or migration), which remain undetermined and are parameterized by the constants $\vec{c}$. The advantage is that these constants can be estimated from single-species experiments. In Fig.~\rev{\ref{fig:inference}}, we show that Eq.~\eqref{eq:model_from_interactions} can be used to infer the coupling parameters ($\alpha$'s in Eq.~\eqref{eq:model_CR_C}) in a consumer-resource model with many species. The errors of the parameter estimation depend on the experiment duration $t$.

We note that the model inferred from Eq.~\eqref{eq:model_from_interactions} is quite general and can, in principle, account for multi-body interactions. This generality arises because the estimated interaction coefficients $\hat{a}_{ij}$ may depend on the densities of all species, denoted by $\vec{x}$. However, the method recovers the classical Generalized Lotka-Volterra model when the coefficients $\hat{a}_{ij}$ are constant:
\begin{equation}\label{eq:Inferred_LV}
    \dot{x}_i = x_i \sum_j \hat{a}_{ij} x_j + x_i c_i.
\end{equation}
Equation~\eqref{eq:Inferred_LV} illustrates the universal emergence of Generalized Lotka–Volterra dynamics from empirical measurements of pairwise interactions, provided these interactions are not sampled across different initial conditions. In other words, when the estimators $\hat{a}_{ij}$ do not depend on species densities—either because densities are sufficiently small for independence to hold approximately, or because interactions are measured only from baseline trajectories with constant initial conditions—the inferred model necessarily reduces to a Generalized Lotka–Volterra form.

\subsection{Interaction measures in stochastic systems}

\begin{figure*}
    \centering
    \includegraphics[width=1\linewidth]{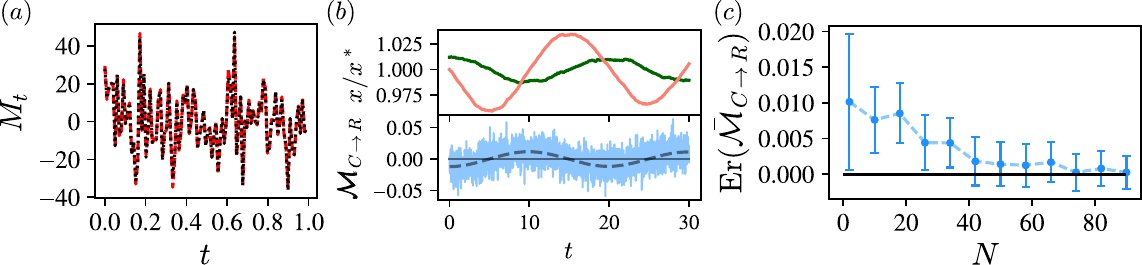}
    \caption{\rev{\textbf{Effect of noise in interaction measures.} In (a), self-interaction measures in a dataset generated with a noisy process (Eq.~\eqref{eq:GBM_model}). In (b, top), trajectory of a consumer and a resource with small white-noise perturbations. In (b, bottom), effect of small noise on interaction estimates in a consumer–resource model from a single realization. The dashed line in (b) indicates the interaction measure in the absence of noise. In (c), absolute difference between the average interaction computed using Eq.~\eqref{eq:average_interaction} and the interaction measured without noise, $\text{Er}(\mathcal{\bar{M}})=|\mathcal{\bar{M}}-\langle\mathcal{M}\rangle|$. As expected, statistical errors decrease as the number of realizations $N$ increases. Synthetic datasets and parameter choices are available in Sec.~\ref{sec:parameters}.}}
    \label{fig:stochastic}
\end{figure*}

\rev{So far, we have used deterministic models to evaluate ecological interactions. Here, instead, we analyze interactions under stochastic dynamics, capturing the intrinsic variability of biological processes. In this setting, the measured interaction matrix becomes a random variable, reflecting variability arising from both the underlying process and the experimental conditions. Even if synthetic data is affected by noise, it is still valid to apply Eqs.~\eqref{eq:def_growth_rate_E} and \eqref{eq:def_interact_E} in the stochastic setting for finite $t$, $\Delta t$, and $\Delta x$. As an illustration, we consider geometric Brownian motion, a minimal model of exponential growth of a small population far from saturation at its carrying capacity,
\begin{equation}\label{eq:GBM_model}
dX_t = v\,X_tdt + D X_t\,dW_t.
\end{equation}
The multiplicative fluctuations in Eq.\eqref{eq:GBM_model} prevent the population from becoming negative and are representative of environmental variability\cite{Grilli2020}. For small $\Delta t$, self-interactions in this model reduce to a simple form (see Appendix~\ref{AP_sec:int_GBM}),
\begin{equation}\label{eq:interactions_GBM_model}
\mathcal{M} (\Delta t)\approx \sqrt{\frac{2}{\Delta t}}\,D\,\mathcal{N}(0,1),
\end{equation}
where $\mathcal{N}(\mu,\sigma^2)$ is a normal random variable with mean $\mu$ and variance $\sigma^2$. Therefore, the self-interaction takes both positive and negative values erratically (Fig.~\ref{fig:stochastic}a).}

\rev{We note that the infinitesimal limit in Eq.~\eqref{eq:limit_empirical_growth_rate} may not exist for time series with abrupt fluctuations. This mathematical remark is relevant to our analysis, since this limit is required for the method in Section~\ref{sec:interactions_close_to_perturbation} to disentangle direct and indirect interactions. For example, time series generated by white-noise-driven processes such as Eq.~\eqref{eq:GBM_model} do not have well-behaved infinitesimal growth rates or self-interactions, as evidenced by the divergence of Eq.~\eqref{eq:interactions_GBM_model} in the limit $\Delta t \to 0$. However, well-behaved empirical growth rate estimates along stochastic paths may still arise, as in synthetic time series generated by stochastic processes with correlated noise—a more natural choice for ecological modeling, as discussed in Appendix~\ref{AP:int_correlated}. Hence, synthetic data aimed at studying interaction measures should be generated using colored-noise stochastic models.} 

\rev{In summary, random fluctuations in population dynamics can cause sign changes in interactions that are unrelated to facilitation–competition trade-offs. Indeed, in Fig.~\ref{fig:stochastic}-b we show that small fluctuations in a simple consumer–resource model considerably increase temporal variability, an effect that can be even more pronounced if stochastic amplification plays a role\cite{McKane2005}. Averaging over $N$ realizations reduces this variability. Specifically, if the same experiment can be repeated $N$ times, each yielding a different empirical interaction matrix $\mathcal{M}^{(k)}{i,j}$, $k=1,\dots,N$, the interaction between species can be estimated as the sample mean over realizations,
\begin{equation}\label{eq:average_interaction}
\bar{\mathcal{M}}_{i,j} = \frac{1}{N}\sum_{k=1}^N \mathcal{M}_{i,j}^{(k)} .
\end{equation}
By virtue of the central limit theorem, this average interaction matrix loses stochasticity as $N$ grows (see Fig.~\ref{fig:stochastic}c). Therefore, while disentangling oscillations induced by population dynamics requires extrapolation to reach the $\Delta t \to 0$ limit, oscillations induced by noise have a fundamentally different nature and can instead be analyzed through extrapolation in $N$ to access the limit $N \to \infty$. In this limit, the average interaction matrix can be interpreted as a statistical correlation,
\begin{equation}
\lim_{N\to\infty}\bar{\mathcal{M}}_{i,j} = \langle\mathcal{M}_{i,j}\rangle = \langle \vec{G}_i \cdot \vec{E}j \rangle,
\end{equation}
highlighting that $\bar{\mathcal{M}}_{i,j}$ expresses a statistical—not necessarily causal—relationship between the growth rate of species $i$ and that of species $j$.}

\section{Discussion}

Measurements of the empirical interaction matrix are influenced by the experimental protocol, particularly the experiment's duration. In short-duration experiments, the interaction matrix directly captures the causal relationships between ecosystem components. However, as the experiment duration increases, these measurements increasingly reflect correlations between per-capita growth rates and density fluctuations, rather than direct causal interactions. This shift does not mean that longer-duration measurements are uninformative. Instead, the emergence of correlations allows the analysis of indirect interactions that may not be evident in direct causal relationships between species~\cite{kefi2012}.

That said, these measurements should be interpreted with caution. For instance, changes in the sign of interactions might be seen as support for the stress-gradient hypothesis, which posits that interactions are a mix of facilitation and competition, with the balance shifting based on environmental context. However, our findings suggest that even when species roles are fixed—meaning there is no explicit dependence on environmental conditions in growth rates and no facilitation-competition trade-offs—the sign of interactions can still change due to intrinsic fluctuations in the system's dynamics. This finding emphasizes that ecologists should consider the possibility of intrinsic temporal fluctuations when interpreting changes in interaction signs observed in empirical data. Moreover, we have shown that measures of interactions can be influenced by experimental decisions, such as how nutrients are supplied to species in controlled growth experiments. Thus, clearly documenting and standardizing experimental protocols is crucial to ensure comparability and reliability of inferred ecological interactions across different studies.

Identifying whether an interaction measure captures direct or indirect species interactions is a challenging task that has traditionally relied on a detailed understanding of the underlying biological processes. Based on such understanding, researchers infer how to interpret the resulting interaction measures~\cite{Paine1966,Paine1974,Wootton1994,Menge1997}. Here, we propose a systematic approach to distinguish the nature of interactions. By evaluating interaction measures at finite times, we can extrapolate their behavior in the short-time limit, which provides direct insight into the directness of the interactions. Practically, this approach offers researchers a robust analytical method for interpreting ecological experiments, even when direct short-term measurements are challenging or noisy.

The interaction matrix does not assume any specific underlying model for its empirical estimation. This raises the natural question of whether such measures can provide insights into plausible generating models for the data, which we address with a proposed inference method. The method is based on sampling short time series over many initial conditions that are then perturbed. This approach offers a paradigm shift compared to typical inference methods, which often require rich longitudinal data—i.e., long and stationary time series. A key feature of the proposed method is that its information is additive: it disentangles single-species and multi-species dynamics, making it possible to separate data from individual species and multi-species experiments, and then combine the information. However, it is important to acknowledge that the accuracy of inferred models is sensitive to the density and precision of data collected at very short timescales, which may be challenging to achieve in field studies.

\rev{The framework developed here demonstrates, within a minimal setting, that measures of interaction may fluctuate even when the relationship between species remains constant over time. Such fluctuations arise from both deterministic population dynamics and stochastic effects, such as measurement error. A first option to address statistical fluctuations is to repeat measurements and replace single estimates with averages, which are less affected by statistical variability. However, performing many experiments may not be feasible in all settings; therefore, robust statistical tools are required to disentangle faithful signals from noise when analyzing single-experiment data~\cite{ives_estimating_2003,klebaner_introduction_2012,aguilar2025limits}. Additional processes, such as delays in species responses and spatial correlations arising from dispersal, may further influence population dynamics and, consequently, the inferred interaction matrices. A systematic analysis of how these processes affect empirical interaction estimates remains an important direction for future work.}

\rev{The main limitation in the study of interactions in ecological communities remains experimental: data are often scarce, sparse, and noisy, and perturbing natural ecosystems in a controlled manner is rarely feasible. These challenges hinder the application and validation of inference techniques. Precisely because of these difficulties, theoretical approaches can play an important role by clarifying the limits of what can be inferred from available data and by helping design experiments that maximize the information gained a priori.}

The main takeaway from this work is  the idea that the experimental protocol is intrinsically linked to the outcomes of measurements, even when assuming error-free data. Therefore, parameters characterizing the experimental setting are just as crucial as those governing species behavior in determining the dynamics of measured interactions. We have shown that mathematical models not only aid in understanding the general properties and underlying mechanisms of ecological systems, but also provide a framework to assess the extent of information in empirical measurements. Ultimately, integrating models with empirical estimates offers a valuable perspective for designing more informative experimental protocols and improving the quality of empirical data.

\section{Methods}

\subsection{Data and code availability statements}\label{sec:data}
Code used to generate synthetic data together with the particular datasets used in figures is available in~\cite{Githubcodes}. Grasshopper's dataset is also present in~\cite{Githubcodes}, but it was extracted from ref.~\cite{Ritchie1993}.

\subsection{Interactions in consumer resource model}\label{sec:measuring_interactions_LRT}

To compute interactions in the general consumer resource model defined in Eqs.~\eqref{eq:model_CR_C} and~\eqref{eq:model_CR_R}, we can use the framework developed in section~\ref{sec:computing_interactions} simply considering a system with $N=n_C+n_R$ entities with densities
\begin{equation}
    x_i =
    \begin{cases}
        C_i,\quad i\le n_C,\\
        R_{i-n_C},\quad i>n_C,
    \end{cases}
\end{equation}
and per-capita growth rates
\begin{equation}\label{eq:growth_rate_general_CR}
    g_i =
    \begin{cases}
        \varepsilon_i\sum_{\ell=1}^{n_R} \alpha_{i,\ell} \,x_{\ell+n_{C}} - d_i,\quad i\le n_C,\\
        \frac{h_i(\vec{x})}{x_i}-\sum_{j=1}^{n_C} x_j\, \alpha_{j,i-n_{C}} ,\quad i>n_C.
    \end{cases}
\end{equation}
An interesting feature of this sort of consumer-resource models is that the gradient vector associated with consumers is time-independent:
\begin{equation}
\vec{G}_i =
\begin{cases}
     \varepsilon_i\sum_{\ell=n_C+1}^{N} \alpha_{i,\ell-n_{C}} \, 
\hat{\ell},\quad i\le n_C, \\
    \nabla\cdot \frac{h_i(\vec{x})}{x_i}-\sum_{j=1}^{n_C}  \alpha_{j,i-n_{C}} \,\hat{j} ,\quad i>n_C.
\end{cases}
\end{equation}
Therefore, any temporal variability in interaction measures associated with consumer growth is driven by the evolution vector $\vec{E}_j$. The behavior of the evolution vector becomes more transparent at the stationary regime, where it can be approximated through linear response theory. Assuming initial conditions in the vicinity of a fixed point $\vec{x}^*$ such that 
\begin{equation}
    \vec{x}^* \, g_i(\vec{x}^*) = 0,\,\forall i,
\end{equation}
the interaction matrix in the small $\Delta t$ and $\Delta x$ limit becomes
\begin{widetext}
\begin{equation}\label{eq:interactions_computed_with_CR_model_linearized}
    \mathcal{M}_{i,j} (\vec{R},t,\Delta x) \frac{1}{\Delta x}\approx
    \begin{cases}
        \varepsilon_i \sum_{k>n_C} \left(e^{\hat{J} t}\right)_{j,k} \alpha_{i,k-n_{C}} ,\quad  i\le n_C, \\
        \frac{d}{d\, R_\ell}\left(\frac{h_\ell(\vec{R})}{R_\ell}\right)\sum_{k>n_C}\left(e^{\hat{J} t}\right)_{j,k}-\sum_{k\le n_C}\left(e^{\hat{J} t}\right)_{j,k}\alpha_{k,i-n_{C}},\, i>n_C,
    \end{cases}
\end{equation}
\end{widetext}
where $\hat{J}$ is the Jacobian,
\begin{equation}
    \left(\hat{J}\right)_{i,j} = \delta_{i,j} \,g_i(\vec{x}^*)+ x_i^*\,\frac{\partial}{\partial_{x_j}} g_i(\vec{x})\Bigg|_{\vec{x}= \vec{x}^*}.
\end{equation}

Eq.~\eqref{eq:interactions_computed_with_CR_model_linearized} was obtained linearizing the dynamics around $\vec{x}^*$, as shown in Appendix~\ref{AP_sec:linearization}. The elements of the exponential matrix appearing in Eq.~\eqref{eq:interactions_computed_with_CR_model_linearized} can be either positive or negative in multi-species models~\cite{Bomze1983,Bomze1995}. Therefore, we expect to observe changes in the sign of interactions measures even in the case of purely competitive ecosystems.

\subsection{Parameters used in figures}\label{sec:parameters}

\begin{itemize}
    \item \underline{Fig.~\ref{fig:data} a,b}: Data generated with yeast-galactose-ethanol model of ref.~\cite{Pacciani-Mori2020},
    \begin{equation}
        \dot{C} = C\left(\sum_{\ell = 1}^2 \nu_\ell \, \alpha_\ell \, \frac{R_\ell}{K_\ell+R_\ell}-d \right),
    \end{equation}
    and
    \begin{equation}
        \dot{R}_\ell = -\,C\left(\alpha_\ell \, \frac{R_\ell}{K_\ell+R_\ell} \right) + C\,\sum_{w=1}^2 \alpha_\ell\,W_{j\ell} \left( \, \frac{R_j}{K_j+R_j} \right) ,
    \end{equation}
    Where $C$ represents the density of yeast, values and parameters with $\ell = 1$ are associated with galactose, and those with $\ell = 2$ with ethanol. Parameter's values correspond to the fit to real data also  performed in ref.~\cite{Pacciani-Mori2020}. $d = 10^{-5}$ $(\text{hours}^{-1})$, $K_1 = 6.73\cdot 10^{-4}$ $(\text{g mL}^{-1})$, $K_1 =  10^{-3}$ $(\text{g mL}^{-1})$, $\nu_1 = 2.03\cdot10^9$ $(\text{g}^{-1})$, $\nu_1 = 4.35\cdot10^{10}$ $(\text{g}^{-1})$ , $\alpha_1 = 1.13\cdot10^{-10}$ $(\text{g hour}^{-1})$, $\alpha_2 = 1.21\cdot10^{-11}$ $(\text{g hour}^{-1})$, $W_{1,1}=W_{2,2}=W_{2,1}=0$, $W_{1,2}=0.3$, $C\left(t=0\right)=162734.08$ $(\text{m}^{-2})$,$R_1\left(t=0\right)=4.5\cdot10^{-3}$ $(\text{g mL}^{-1})$, $R_2\left(t=0\right)=0$. Regarding the sampling times and perturbations, $\Delta t= 1$ (hour), $\Delta x_R=10^{-4}$ $(\text{g mL}^{-1})$, $\Delta x_C=10^{4} \,(\text{m}^{-2})$. \rev{Uncertainties are not available since they are not provided in Ref.~\cite{Pacciani-Mori2020}. The parameters are used to reproduce realistic temporal scales of the system rather than for detailed parameter estimation.}
    
    \item \underline{Fig.~\ref{fig:interactions_1C1R}}: Data generated using model in  Eq.~\eqref{eq:1C_1R_CR_R_eq}. In a-d, $n_C=1$, $n_R=1$, $a=1.2$, $K=2$, $d=0.1$, $r=1$. In b,c, and d, respectively,  $\varepsilon=5\cdot10^{-2}$, $\varepsilon=0.12$, and $\varepsilon=10$. Initial conditions coincide with stable fixed points (Eq.~\eqref{eq:fx_points_1C_1R}) 
    \item \underline{Fig.~\ref{fig:competitive_systems}}: In a-c, data generated using Eq.~\eqref{eq:model_CR_C}  with $n_C=2$, $n_R=2$,  $\alpha_{1,1} = 0.05$, $\alpha_{1,2} = 1$ , $\alpha_{2,2} = 2$,  $\alpha_{2,1} = 0$, $K=2$, $d=0.1$, $r=3$, $R(t=0)=C(t=0)=1$, $\varepsilon = 0.08$ in e, and $\varepsilon = 0.5$ in f. 

    \item \underline{Fig.~\ref{fig:scaling}}: Data generated using model in Eqs.~\eqref{eq:model_CR_C} and~\eqref{eq:model_CR_R}. In b, $n_R=n_C=10$, renewal function \makebox{$h_\ell(\vec{R}) = \Gamma_\ell = 100$}. Defining $U_{[a,b]}$ as an uniformly distributed random variable in the interval $[a,b]$, $d\sim U_{[1,10]}$, $\tilde{\alpha}\sim U_{[0,1]}$, $\alpha_{i,j} =\tilde{\alpha}_{i,j}d_i/\sum_m\tilde{\alpha}_{i,m} $,  $K\sim U_{[1,10]}$, $C(t=0)\sim U_{[1,10]}$, $R(t=0)\sim U_{[1,10]}$. In c, same as in Fig.~\ref{fig:competitive_systems}-b.
    
    \item \underline{Fig.~\ref{fig:experiment_influence}}: Data generated using model in Eqs.~\eqref{eq:model_CR_C} and~\eqref{eq:model_CR_R} with $n_C=n_R=2$,  $r=1$, $\varepsilon=0.1$, $\alpha\sim U_{[0,1]}$, $K=5$, $\Gamma=1$, $C(t=0)\sim U_{0,10^{-3}}$. 
    
    \item \underline{Fig.~\ref{fig:inference}}: $n_C=n_R=20$, $\alpha\sim U_{[0,2]}$.

    \item \rev{\underline{Fig.~\ref{fig:stochastic}}: In a, data generated with Eq.~\eqref{eq:GBM_model} with $v=1$, $D=1$, $X_0=1$. In b and c, same as in Fig.~\ref{fig:interactions_1C1R} d, with additive noise and diffusion constant $D_C=D_R=10^{-4}$.}
\end{itemize}

\section{Acknowledgments}
We thank Sara Mitri, Davide Bernardi and Luca Martinoia for their useful comments. J.A., S.A. and A.M. acknowledge financial support under the National Recovery and Resilience Plan (NRRP), Mission 4, Component 2, CUP
2022WPHMXK, Investment 1.1, funded by the European Union – NextGenerationEU – Project Title:  ``Emergent Dynamical Patterns of Disordered
Systems with Applications to Natural Communities” . S.S. acknowledges financial support under the National Recovery and Resilience Plan (NRRP), Mission 4, Component 2, Investment 1.1, Call for tender No. 104 published on 2.2.2022 by the Italian Ministry of University and Research (MUR), funded by the European Union – NextGenerationEU – Project Title: Anchialos: diversity, function, and resilience of Italian coastal aquifers upon global climatic changes – CUP C53D23003420001 Grant Assignment Decree n. 1015 adopted on 07/07/2023 by the Italian Ministry of Ministry of University and Research (MUR).

\section{Author Contributions}
JA, SS, AM, and SA conceived and designed the study. JA performed the calculations, simulations, and data analysis. JA, SS, AM, and SA wrote the manuscript, and all authors reviewed and approved the final version.

\section{Competing Interests statement}
There are no competing interests.

\clearpage
\newpage

\onecolumngrid
\bibliography{refs}% Produces the bibliography via BibTeX.

\clearpage

\appendix
\renewcommand{\appendixname}{Appendix}
\renewcommand{\thesection}{\Alph{section}}
\renewcommand{\thesubsection}{\Alph{section}.\arabic{subsection}}

% Redefine figure numbering in supplementary
\renewcommand{\thefigure}{A~\thesection\arabic{figure}}
\makeatletter
\@addtoreset{figure}{section} % reset figure counter at each new section (A, B, …)
\makeatother

\section{Measures of interactions in Grasshoppers data}\label{sec:grasshoppers}

\rev{In Ref.~\cite{Ritchie1993}, time series of abundances are reported for three grasshopper species: Melanoplus femur-rubrum, Spharagemon collare, and Phoetaliotes nebrascensis. The time series span approximately one month and correspond to experiments in which grasshoppers were maintained under different initial abundances and species combinations. In particular, the dataset includes time series for each species growing alone. We use these single-species trajectories as baseline dynamics to compute interactions. The dataset also includes time series for each pairwise combination of species, which we treat as perturbed trajectories.}

\rev{For example, consider the estimate of the interaction $\mathcal{M}_{S\to M}$, this is, how adding Spharagemon collare (noted S) affects the growth of  Melanoplus femur-rubrum (noted M). The time series  corresponding to the evolution of M with no other grasshopper species is used as baseline trajectory. In such trajectory, empirical growth rates are measures through Eq.~\eqref{eq:def_growth_rate_E}. Similarly, empirical growth rates of M are measured in an experiment in which S was also present. Such growth rates corresponds to the perturbed evolution. Once both growth rates are computed, the interaction is obtained through Eq.~\eqref{eq:def_interact_E}.}

\begin{figure}
    \centering
    \includegraphics[width=0.9\linewidth]{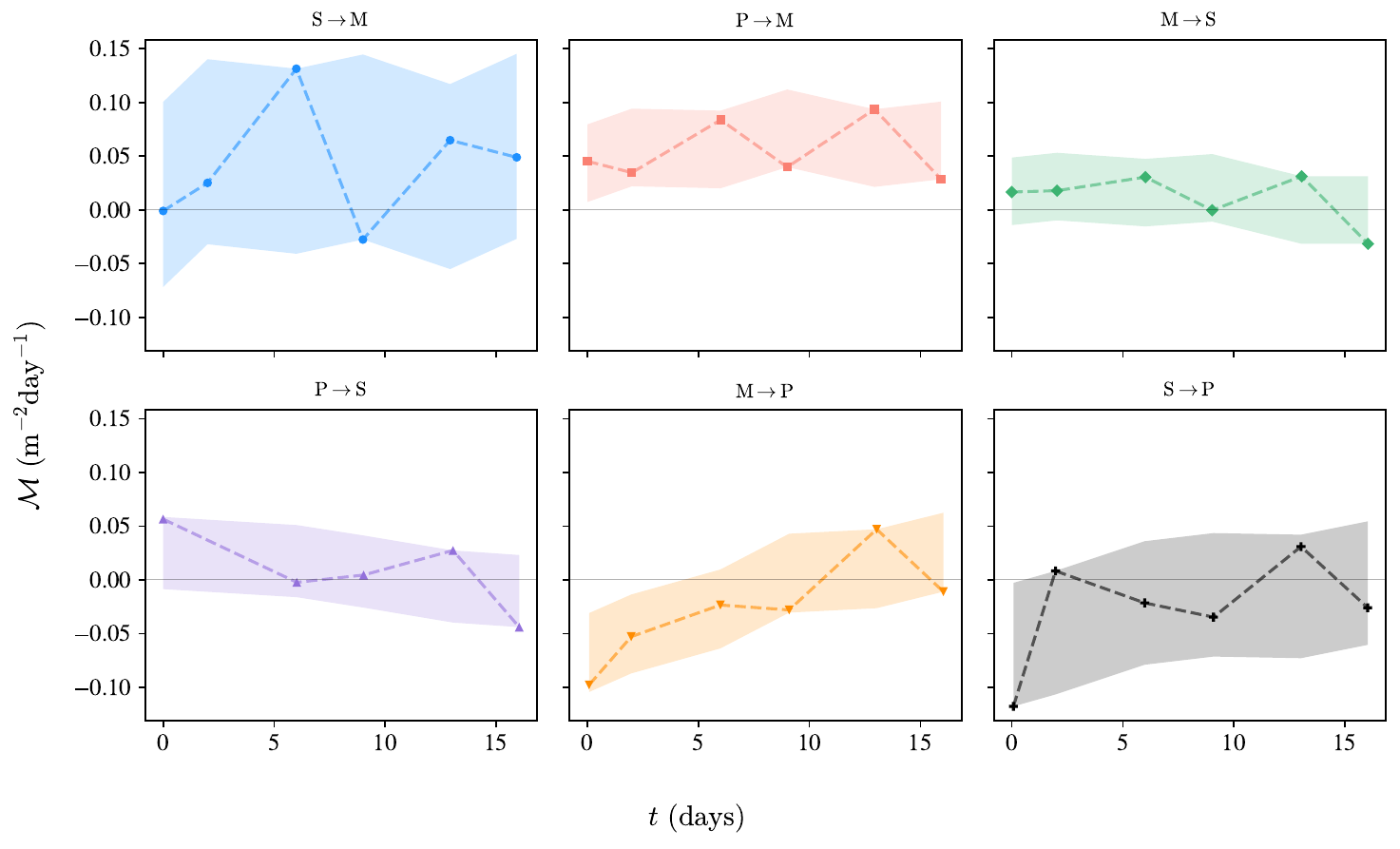}
    \caption{\rev{Estimated interspecific interaction time series for the grasshopper dataset. Each panel shows the interaction measure between a pair of species, computed as the difference between per-capita growth rates in the perturbed (pairwise) and baseline (single-species) trajectories. Points represent the estimated interactions at each time step. Shaded bands correspond to $95\%$ pointwise confidence intervals obtained using a residual bootstrap procedure with 1000 resamples. The intervals reflect sampling uncertainty in the estimated growth-rate increments. The horizontal line at zero indicates the absence of interaction.}}
    \label{fig:confidence_intervals}
\end{figure}

\rev{In Fig.~\ref{fig:confidence_intervals}, we report $95\%$ pointwise confidence intervals for the estimated interactions. These intervals were obtained using a residual bootstrap procedure applied to the growth-rate increments. For each bootstrap sample, the increments were resampled with replacement, the interaction time series was recomputed, and confidence bands were constructed from the empirical $2.5$ and $97.5$ percentiles of the bootstrap distribution over $1000$ realizations.}

\section{Analytical expression for interactions evaluated on synthetic data}\label{AP_sec:proof_formula_interactions}

In this section, we provide the proof for Eq.~\eqref{eq:interactions_computed_with_the_model} from the main text, which enables the evaluation of the empirical interaction matrix when applied to a synthetic dataset generated by a model with a known per-capita growth rate. The derivation makes use of multidimensional Taylor expansions. Consider that the per-capita growth rate is evaluated on a vectorial function $\vec{F}$ that depends itself on a vectorial quantity $\vec{y}$ : $g_i[\vec{F}(\vec{y})]$. Then, let us perturb the $j^{th}$ component of $y$ in an small amount $\Delta y$,
\begin{equation}
    g_i\left[\vec{F}(\vec{y}+\Delta y \,\hat{j})\right] = g_i\left[\vec{F}(\vec{y})+\hat{j}\cdot\nabla\vec{F}(\vec{y})\, \Delta y +\mathcal{O}(\Delta y^2)\right] = g_i\left[\vec{F}(\vec{y})\right]+\nabla  g_i\left[\vec{F}(\vec{y})\right]\cdot\partial_{y_j}\vec{F}(\vec{y}) \,\Delta y+\mathcal{O}(\Delta y^2).
\end{equation}
Therefore,
\begin{equation}
    g_i\left[\vec{F}(\vec{y}+\Delta y \,\hat{j})\right]-g_i\left[\vec{F}(\vec{y})\right] = \nabla  g_i\left[\vec{F}(\vec{y})\right]\cdot\partial_{y_j}\vec{F}(\vec{y}) \,\Delta y+\mathcal{O}(\Delta y^2).
\end{equation}
The above expression directly leads to Eq.~\eqref{eq:interactions_computed_with_the_model} by substituting $\vec{y}$ by $\vec{x}^{(0)}$, and $F(\vec{y})$ by $\vec{x}(t|\vec{y})$.

\section{Interactions in single species systems -- self interactions}\label{AP:One_dim_int}

As discussed in section~\ref{sec:computing_interactions} of the main text, it is not generally possible to compute the value of empirical interactions measured over synthetic data as this entails solving the entire dynamics of the species densities. However, this is feasible in one dimensional systems, where one would obtain $x(t|x^{(0)})$ solving the equation
\begin{equation}\label{eq:one_dimensional_solution}
    \int_{x^{(0)}}^{x(t|x^{(0)})} \frac{ds}{s g(s)}=t.
\end{equation}
Moreover, it is possible to extract information about interaction measures even without solving Eq.~\eqref{eq:one_dimensional_solution}.
Let us re-write eq.~\eqref{eq:def_model} in a one-dimensional setting and making explicit the dependence on the initial condition
\begin{equation}
    \frac{\partial}{\partial t}x(t|x^{(0)})= x(t|x^{(0)}) g\left(x(t|x^{(0)})\right),
\end{equation}
deriving with respect to  $x^{(0)}$ the above equation and assuming that derivatives can be exchanged, we can write an ordinary differential equation for $ \frac{\partial}{\partial x^{(0)}}x(t|x^{(0)})$,
\begin{align}
    \frac{\partial}{\partial t}  \frac{\partial}{\partial x^{(0)}}   x(t|x^{(0)})= H\left[x(t|x^{(0)})\right]\frac{\partial}{\partial x^{(0)}}x(t|x^{(0)}),
\end{align}
with
\begin{equation}
    H\left[x(t|x^{(0)})\right] =  g\left(x(t|x^{(0)})\right)+x(t|x^{(0)}) g'\left(x(t|x^{(0)})\right).
\end{equation}
Therefore,
\begin{align}
    \frac{\partial}{\partial x^{(0)}} & x(t|x^{(0)}) = \exp\left[\int_0^t ds \, H\left[x(s|x^{(0)})\right]\right]
\end{align}
where we fixed the integrating constant imposing $  \frac{\partial}{\partial x^{(0)}}x(0|x^{(0)})=1$. Using this result we can rewrite eq.~\eqref{eq:interactions_computed_with_the_model} as
\begin{equation}
    \mathcal{M}  (t,\Delta x,\Delta t) =    g'(x(t|x^{(0)})) \exp\left[\int_0^t ds \, H\left[x(s|x^{(0)})\right]\right] \Delta x  
+\mathcal{O}(\Delta t)+\mathcal{O}(\Delta x^2).
\end{equation}
Since the range of $H(\cdot)$ belongs to the reals, the exponential in the above equation is always positive and changes in the sign of the interactions only depend on the sign of the derivative of the per-capita growing rate. Therefore, in this setting changes of sign in the self-interactions only can arise from competition-facilitation trade-offs.

\subsection{Slef interactions with logistic growth}\label{sec:logistic}

To illustrate measures of self-interactions in single-species growth, we analyze the logistic growth:
\begin{equation}
    \dot{x} = r\, x\left(1-\frac{x}{K}\right),
\end{equation}
with the solution
\begin{equation}\label{eq:sol_logistic_growth}
    x(t|x_0) = \frac{x_0\, K\, e^{-r\,t}}{K - x_0 + x_0\, e^{-r\,t}}.
\end{equation}
Eq.~\eqref{eq:sol_logistic_growth} can be used to generate time series with the desired sampling time $\Delta t$, magnitude of perturbation $\Delta x$, and experiment duration $t$. Moreover, having the explicit solution allows us to evaluate Eq.~\eqref{eq:interactions_computed_with_the_model} exactly. In the case of single-species evolution, the interaction matrix reduces to a single value that accounts for self-interactions — i.e., the effect of small perturbations in the species population on the overall density growth.

\begin{figure*}
    \centering
    \includegraphics[scale=0.8]{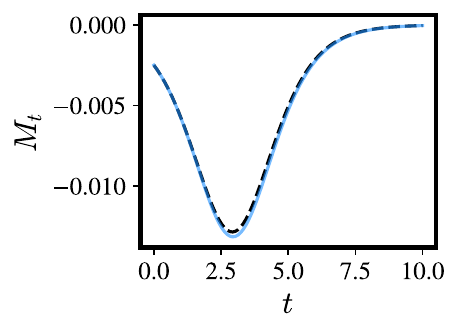}
    \caption{Self-interactions in the logistic model. Interactions computed on synthetic data generated with Eq.~\eqref{eq:sol_logistic_growth} (continuous blue lines) and using the analytical approximations results (dark dashed lines) from eq.~\eqref{eq:interactions_computed_with_the_model}.}
    \label{fig:logistic_interactions}
\end{figure*}

In Fig.~\ref{fig:logistic_interactions} we show that  Eq.~\eqref{eq:interactions_computed_with_the_model} properly describe the self-interactions measured from realizations of the process. The sign of self-interactions is always negative, being consistent with the system's dynamics, as increasing the density only decelerates the per-capita growth if $x^{(0)}<K$ or accelerates the per-capita decay if $x^{(0)}>K$. We also observe that interactions converge to zero as the experiment duration $t$ grows, this trend is expected as both the baseline and perturbed trajectories will end up at the same stable fixed point $x^*=K$. Indeed, in the logistic growth $g'(x)=-r/K$, and thus changes of sign of self-interactions are not possible.

%Therefore, changes of sign in self-interactions in single-species systems are only observed as the result of competition-facilitation trade-offs, i.e. when the gradient of the per-capita growth rate has two contributions $g'(x)=A(x)+B(x)$ where $A>0$ and $B<0$. Competition-facilitation trade-offs in single-species has been observed empirically~\cite{Meacock2025} and according to our derivation they align with the stress-gradient hypothesis. However, as we shall see in the next section, multi-species systems can exhibit changes of sign in interaction even in the absence of competition-facilitation trade-offs.

\section{Interaction measures close to a fixed point}\label{AP_sec:linearization}
In this section, we consider initial conditions in the vicinity of a fixed point $\vec{x}^*$ such that 
\begin{equation}
    \vec{x}^* \, g_i(\vec{x}^*) = 0,\forall i.
\end{equation}
When $\vec{x}^{(0)}= \vec{x}^*$,  the equation of motion for the evolution of the perturbed time series can be linearized. Considering
\begin{equation}
    \delta\vec{x}(t,j) = \vec{x}(t|\vec{x}^*+\Delta x\,\hat{j})-\vec{x}^*,
\end{equation}
then
\begin{equation}
    \dot{\delta\vec{x}} = \hat{J}\delta\vec{x},
\end{equation}
where $\hat{J}$ is the Jacobian,
\begin{equation}
    \left(\hat{J}\right)_{i,j} = \delta_{i,j} \,g_i(\vec{x}^*)+ x_i^*\,\frac{\partial}{\partial_{x_j}} g_i(\vec{x})\Bigg|_{\vec{x}= \vec{x}^*}.
\end{equation}
Therefore, trajectories in the vicinity of fixed points looks like
\begin{equation}
    \vec{x}(t|\vec{x}^*+\Delta x\,\hat{j})=  \vec{x}^*+e^{\hat{J} t}\delta\vec{x}(0,j) =  \vec{x}^*+e^{\hat{J} t}\left(\vec{x}^*+\Delta x\,\hat{j}\right).
\end{equation}
The derivative of the above expression with respect to $\Delta x$ gives the Evolution vector, which we can identify as the $j-$th column of the exponential matrix $e^{\hat{J}t}$,
\begin{equation}
    \vec{E}_j=\sum_k \hat{k} \, \left(e^{\hat{J} t}\right)_{j,k} .
\end{equation}
Therefore, interactions in the vicinity of fixed points can be computed as
\begin{align}\label{eq:interactions_computed_with_the_model_linearized}
    \mathcal{M}_{i,j}  =& \sum_k \left(e^{\hat{J} t}\right)_{j,k} \left(\vec{G}_i\right)_{k} \,   \Delta x +\mathcal{O}(\Delta t)+\mathcal{O}(\Delta x^2).
\end{align}

\section{Linear analysis in one consumer-one resource model}\label{AP_sec:1C1R}

In this section, we make the linear stability analysis for the simple model
\begin{equation}
    \dot{C} =\varepsilon \alpha C  R - C d,
\end{equation}
and
\begin{equation}
    \dot{R} = r R \left(1-\frac{R}{K}\right)-   \alpha R C.
\end{equation}
Among the model's five parameters—$r$, $K$, $\alpha$, $d$, and $\varepsilon$—three can be eliminated through rescaling of the variables $R$, $C$, and $t$. This leaves only two independent parameters that govern the possible behaviors of the interactions. For our analysis, we choose these to be $r$ and, in particular, $\varepsilon$.

The eigenvalues of the Jacobian of the dynamics evaluated at the coexistence fixed point, 
\begin{align}\label{eq:fx_points_1C_1R}
    R^* &=\frac{d}{\alpha\varepsilon},\nonumber   \\
    C^* &= \frac{r}{\alpha}\left(1-\frac{R^*}{K}\right),
\end{align}
reads
\rev{
\begin{equation}
    \lambda_\pm = -\lambda \,(1\pm \sqrt{\omega}),\quad\text{with} \quad\lambda=\frac{d r}{2\alpha K\varepsilon}, \quad\text{and} \quad \omega= 1+\frac{4\alpha K\varepsilon}{r}-\frac{(2 \alpha K\varepsilon)^2}{rd}.
\end{equation}
}
The coexistence fixed point is always stable, as the real parts of both eigenvalues are negative. However, this fixed point has positive components only if \( R^* < K \), or equivalently, if \( \varepsilon < \frac{d}{\alpha K} \). Therefore, we focus on this parameter region.

The time scales of damping and oscillations characterizing the relaxation towards equilibrium are related with the real and imaginary parts of the eigenvalues respectively   
\begin{equation}
    \tau_D^{-1} = \left| \text{Re}(\lambda_+) \right|=
    \begin{cases}
        \lambda,\quad\text{if}\quad \varepsilon > \varepsilon_{\text{TH}}, \\
        \left|\lambda_+\right|,\quad\text{if}\quad \varepsilon \le \varepsilon_{\text{TH}}.
    \end{cases}
\end{equation}
and the oscillation timescale,
\begin{equation}
    \tau_O^{-1} = \left| \text{Im}(\lambda_+) \right|=
    \begin{cases}
        \lambda\sqrt{\left|w\right|},\quad\text{if}\quad \varepsilon > \varepsilon_{\text{TH}}, \\
        0,\quad\text{if}\quad \varepsilon \le \varepsilon_{\text{TH}}.
    \end{cases}
\end{equation}
with $\varepsilon_{\text{TH}} = \frac{\sqrt{dr + d^2} + d}{2aK}$.

$\tau_D$ characterizes the exponential decay toward the coexistence fixed point, with larger $\tau_D$ indicating a slower decay. Similarly, $\tau_O$ determines the frequency of oscillations, with larger $\tau_O$ corresponding to greater temporal distance between consecutive peaks. The ratio
\begin{equation}
    \frac{\tau_D}{\tau_O} =
    \begin{cases}
        \sqrt{\left|w\right|}, & \text{if } \varepsilon > \varepsilon_{\text{TH}}, \\
        0, & \text{if } \varepsilon \le \varepsilon_{\text{TH}},
    \end{cases}
\end{equation}
describes the relative importance of damping versus oscillatory behavior. A small value of $\tau_D/\tau_O$ indicates that oscillations are slower than damping, and thus perturbation relaxation is expected to be mostly monotonic. In contrast, a large $\tau_D/\tau_O$ corresponds to a highly fluctuating relaxation, as illustrated in Fig.~\ref{fig:competitive_systems}.

The oscillation threshold $\varepsilon_{\text{TH}}$ increases with $r$. Therefore, as $r$ increases and $\varepsilon$ decreases, oscillations become less significant. In this limit, the resource evolves much faster than the consumer, allowing the consumer dynamics to be effectively decoupled from the resource. The resulting approximation is:
\[
\dot{C} \approx \tilde{r} \, C \left(1 - \frac{C}{\tilde{K}} \right),
\]
where $\tilde{r} = \varepsilon a K - d$, and $\tilde{K} = \frac{r(\varepsilon a K - d)}{\varepsilon a^2 K}$. Therefore, in this time-scale separation limit, oscillations are suppressed as the system effectively lives in a one-dimensional space.

\section{Expansion of the empirical interaction matrix for short experiment durations}\label{AP_sec:Expansion_experiment_duration}

In this section, we provide more details about the derivation of the equations apearing in section~\ref{sec:interactions_close_to_perturbation}. Integrating the model definition we obtain,
\begin{equation}\label{eq:integration}
    x_k(t|\vec{y}) = y_k+\int_0^t ds\,g_k\left[\vec{x}(s|\vec{y})\right]\,x_k(s|\vec{y}).
\end{equation}
Where $\vec{y}=\vec{x}^{(0)}$.

The Taylor series expansion to first order reads
\begin{equation}
    x_k(t|\vec{y}) = y_k + v_k \, t + \mathcal{O}(t^2),
\end{equation}
were we defined the `velocity' 
\begin{equation}
    \vec{v}=\sum_\ell y_\ell \, g_\ell\left(\vec{y}\right)\hat{\ell}.
\end{equation}
Inserting the above series-expansion again in Eq.~\eqref{eq:integration}, we increase one order in the expansion
\begin{align}
    x_k(t|\vec{y}) &= y_k+\int_0^tds\, g_k\left[\vec{y} + \vec{v} \,s + \mathcal{O}(s^2)\right]\left[y_k + v_k \, s + \mathcal{O}(s^2)\right]= \nonumber \\
    &= y_k+\int_0^tds\,\left[g_k(\vec{y})+\nabla_{\vec{y}} \, g_k(\vec{y}) \cdot \vec{v}\,s+ \mathcal{O}(s^2)\right]\left[y_k + v_k\, s + \mathcal{O}(s^2)\right]= \nonumber \\
    &=y_k+v_k\, t + \left[v_k \,g_k(\vec{y})+y_k\nabla_{\vec{y}} \, g_k(\vec{y}) \cdot \vec{v}\right] \,\frac{t^2}{2}+\mathcal{O}(t^3).
\end{align}
Using 
\begin{equation}
    \nabla \left(v_k\right) = \nabla \left(y_k \,g_k\right) = y_k \, \nabla g_k + g_k\hat{k},
\end{equation}
then 
\begin{equation}\label{eq:evolution_x_second_order}
x_k(t|\vec{y}) = y_k +v_k\,t+\nabla_{\vec{y}}\,v_k\cdot\vec{v}\,\frac{t^2}{2}+\mathcal{O}(t^3).
\end{equation}
which leads to the following expression for the components of the evolution vector:
\begin{equation}\label{eq:approx_evolution}
   \left(\vec{E_j}\right)_k =\frac{\partial}{\partial_{y_j}} \,x_k(t|\vec{y})  = \delta_{j,k} + \partial_{y_j}\,v_k\, t + \partial_{y_j}\left[\nabla_{\vec{y}}v_k\cdot\vec{v}\right]\,\frac{t^2}{2}+\mathcal{O}(t^3).
\end{equation}

Next, using eq.~\eqref{eq:evolution_x_second_order} we approximate the gradient to second order:
\begin{align}\label{eq:approx_gradient_first_order}
     \vec{G}_i & =\nabla\,  g_i\left[\vec{x}(t|y)\right] = \nabla\,  g_i\left[\vec{y} +\vec{v}\,t+\sum_\ell\left(\nabla\,v_\ell\cdot\vec{v}\right)\hat{\ell}\,\frac{t^2}{2}+\mathcal{O}(t^3)\right]=
     \nonumber \\
     &=\nabla \,g_i(\vec{y}) +  \sum_{k} \hat{k}\, \nabla \left[\partial_{y_k} g_i(\vec{y})\right]\,\cdot  \left[\vec{v}\,t+\sum_\ell\left(\nabla\,v_\ell\cdot\vec{v}\right)\hat{\ell}\,\frac{t^2}{2}\right]  + \mathcal{O}(t^3),
\end{align}
Inserting Eqs.~\eqref{eq:approx_evolution} and~\eqref{eq:approx_gradient_first_order} in Eq.~\eqref{eq:interactions_computed_with_the_model} we can get the coefficients of the interaction matrix,
\begin{equation}
    \mathcal{M}_{i,j}=\mathcal{M}^{(0)}_{i,j}+\mathcal{M}^{(1)}_{i,j}\,t+\mathcal{M}^{(2)}_{i,j}\,t^2+\dots
\end{equation}
In particular,
\begin{equation}
    \frac{\mathcal{M}^{(0)}_{i,j}}{\Delta x} = \partial_{y_j}g_i(\vec{y}),
\end{equation}
and
\begin{equation}\label{eq:M_1}
    \frac{\mathcal{M}^{(1)}_{i,j}}{\Delta x} = \nabla\left(\partial_{y_j}g_i(\vec{y})\right)\cdot\vec{v}\,+\,\nabla g_i(\vec{y})\cdot\partial_{y_j}\vec{v}.
\end{equation}

%and
%\begin{equation}
%    \mathcal{M}^{(2)}_{i,j} = \frac{1}{2}\partial_{y_j} \left[\nabla g_i\cdot\sum_\ell\left(\nabla v_\ell\cdot\vec{v}\right)\,\hat{\ell}\right]+\sum_k\partial_{y_j}v_k \left[\nabla \left(\partial_{y_k} \,g_i\right)\cdot \vec{v}\right].
%\end{equation}

\subsection{Expansion of interaction matrix in environmental-organism models}\label{AP_sec:OE_model}

Environmental-organisms models are a generalization of consumer-resource models proposed in ref.~\cite{Meacock2025}  to capture a broader range of interactions between species and environmental factors.  In particular, the model is defined as follows
\begin{equation}\label{eq:EO_model_C_1R1C}
    \dot{C}_i = C_i \,g_i\left(\vec{R}\right),
\end{equation}
and
\begin{equation}\label{eq:EO_model_R_1R1C}
    \dot{\vec{R}}_\ell = \sum_\ell C_\ell \vec{f}_\ell\left(\vec{R}\right)+h_\ell(\vec{R}).
\end{equation}
Here, the vector $\vec{R}$ represents environmental traits that influence species growth, including resource availability, temperature, light, and other abiotic factors~\cite{Meacock2025}.

Notably, this model does not include direct interactions between species populations ($C_i$). As a result, the leading-order term in the time-expansion of the interaction measure vanishes:

\begin{equation}
    \mathcal{M}^{(0)}_{i,j} = 0.
\end{equation}

Thus, the first non-zero contribution is the next-order term $\mathcal{M}^{(1)}_{i,j}$, which captures fast, indirect interactions mediated by the environment. The expression for $\mathcal{M}^{(1)}_{i,j}$ in Eq.~\eqref{eq:M_1} simplifies when both indices refer to species since:
\begin{equation}
    \nabla_{\vec{x}} \,g_i \left(\vec{R}\right) = \nabla_{\vec{R}} \,g_i \left(\vec{R}\right),
\end{equation}
\begin{equation}
    \partial_{C_j}\, \vec{v} = g_j\left(\vec{R}\right)\,\hat{j}+\vec{\tilde{f}}_j,
\end{equation}
and
\begin{equation}
    \partial_{C_j} g_i\left(\vec{R}\right)=0,
\end{equation}
where $\nabla_{\vec{x}}$ and $\nabla_{\vec{R}}$ represent the nabla operator in the space of species and environmental vector jointly and the nabla operator in the space of environmental vector respectively. We also defined $\vec{\tilde{f}}_j$ as a vector that has zeros in the species components and is equal to $\vec{f}_j$ in the environmental components. 

Substituting these expressions into Eq.~\eqref{eq:M_1}, we obtain the environmental-mediated interaction:
\begin{equation}\label{eq:Mitris_work}
    \frac{\mathcal{M}^{(1)}_{i,j}}{\Delta x} = \nabla_{\vec{R}} \,g_i \left(\vec{R}\right) \cdot \vec{f}_j,
\end{equation}
which corresponds to the instantaneous interaction matrix defined in Ref.~\cite{Meacock2025}.

\subsection{Expansion of interaction matrix in consumer-resource models}\label{AP_sec:CR_model}

In this section, we give explicit form for the expansion of the interaction matrix in the consumer-resource model defined in Eqs.~\eqref{eq:model_CR_C} and ~\eqref{eq:model_CR_R}. In particular, we focus in the interaction between two consumers. The gradient vector associated to consumers read
\begin{equation}\label{eq:gradient_vector_CR}
\vec{G}i = \varepsilon_i\sum_{\ell=n_C+1}^{N} \alpha_{i,\ell-n_{C}} \, 
\hat{\ell},\quad i\le n_C.
\end{equation}
The coefficients $\mathcal{M}^{(0)}$, $\mathcal{M}^{(1)}$, and $\mathcal{M}^{(2)}$ are obtained inserting Eqs.~\eqref{eq:approx_evolution} and~\eqref{eq:gradient_vector_CR} in Eq.~\eqref{eq:interactions_computed_with_the_model}:
\begin{equation}
    \mathcal{M}^{(0)}_{i,j}=\varepsilon_i\sum_{\ell>n_C}\alpha_{i,\ell-n_C} \,\delta_{j,\ell} = 0,
\end{equation}
\begin{equation}
    \mathcal{M}^{(1)}_{i,j}=\varepsilon_i\sum_{\ell=1}^{n_R}\alpha_{i,\ell} \,\partial_{C_j}\left(R_k \,g_{_{R_k}}\right)=-\varepsilon_i\sum_{k=1}^{n_R}\alpha_{i,k}\,\alpha_{j,k} \,R_k,
\end{equation}
and
\begin{align}
    \mathcal{M}^{(2)}_{i,j}&=\frac{\varepsilon_i}{2}\sum_{k=1}^{n_R}\alpha_{i,k} \,\partial_{C_j}\left(\nabla_{\vec{}x}\, v_{k+n_C}\cdot\vec{v}\right)  \nonumber \\ 
    &=\frac{\varepsilon_i}{2}\sum_{\ell=1}^{n_R}\alpha_{i,\ell} \,\left[-\alpha_{j,j}R_k\left(g_j+g_{k+n_C}\right)+\sum_{\ell=1}^{n_R} \partial_{R_\ell} h_\ell\,\alpha_{j,\ell}\,R_\ell+\sum_{w<n_C}\alpha_{w,k}\,\alpha_{j,k}R_k\right].
\end{align}

\section{Model inference from interaction matrix}\label{AP_sec:inferring_model_from_interactions}

If we were to measure the empirical interaction matrix, $\mathcal{M}_{i,j} (\vec{x},t,\Delta x,\Delta t)$, based on synthetic paths generated by the general model in Eq.\eqref{eq:def_model} for small values of $\Delta t$, $\Delta x$, and $t$, we could access the limit
\begin{equation}\label{eq:def_aij}
    \lim_{\substack{t\to 0 \\ \Delta x \to 0 \\ \Delta t\to 0}} \frac{1}{\Delta x} \mathcal{M}_{i,j} (\vec{x},t,\Delta x,\Delta t) =a_{ij} \left(\vec{x}\right), 
\end{equation}
where
\begin{equation}
      a_{ij} \left(\vec{x}\right) = \frac{\mathcal{M}_{i,j}^{(0)} (\vec{x}) }{\Delta x}
\end{equation}
is the theoretical definition of the interaction matrix~\cite{Novak2016}. We can define the following vector function
\begin{equation}
    \vec{F}_i(\vec{x}) = \sum_j a_{ij}  \left(\vec{x}\right) \hat{j},
\end{equation}
which, in virtue of Eq.~\eqref{eq:zeroth-order_approx_M} fulfills,
\begin{equation}\label{eq:exact_differential}
    \vec{F}_i(\vec{x}) \cdot d\vec{x} = \left(\nabla \cdot g_i \right) \cdot d\vec{x} = d g_i .
\end{equation}
Therefore, the vector field $\vec{F}_i(\vec{x})$ is conservative, i.e.  $\vec{F}_i$ derives from a potential (scalar field) that we can identify as the per-capita growth rate of species $i$.

Now let us consider the situation in which we have real data with finite $t$, $\Delta t$, and $\Delta x$. From these data, we could compute the estimators
\begin{equation}
    \hat{a}_{ij} = \frac{1}{\Delta x} \mathcal{M}_{i,j} (\vec{x},t,\Delta x,\Delta t),
\end{equation}
and 
\begin{equation}
    \hat{\vec{F}}_i(\vec{x}) = \sum_j \hat{a}_{ij}  \left(\vec{x}\right) \hat{j}.
\end{equation}

Then if $\hat{\vec{F}}_i(\vec{x})$ is conservative, i.e. if $\nabla \cdot \hat{\vec{F}}_i\approx 0 $, then Eq.~\eqref{eq:exact_differential} can be seen as a differential equation for the per-capita growth rate with solution,
\begin{equation}
    \hat{g}_i\left(\vec{x}\right) =  \int_0^{t} \sum_j  \hat{a}_{ij} \left(s\right)  ds + \hat{g}_i\left(\vec{0}\right).
\end{equation}
Moreover, considering again that $\hat{\vec{F}}_i$ is conservative, then its integral does not depend on the integration path, and we can write 
\begin{equation}\label{eq:inference_of_growth_rate}
    \hat{g}_i\left(\vec{x}\right) = c_i+  \sum_j \int_0^{x_j} \hat{a}_{ij} \left(x_1,\dots,x_{j-1},s,x_{j+1},\dots\right)  ds,
\end{equation}
where $c_i =\vec{g}\left(\vec{0}\right)$ are  integration constants, independent of the measure of interactions.

Using the definition of the local growth rate (Eq.\eqref{eq:def_model}), the model inferred from the leading term of interactions would read
\begin{equation}
    \dot{x}_i = x_i \sum_j \int^{x_j}_0 \hat{a}_{ij} \left(x_1,\dots,x_{j-1},s,x_{j+1},\dots\right)  ds_j+x_i c_i.
\end{equation}

We note that the inference framework relies on the assumption that the empirical vector field, $\hat{\vec{F}}_i = \sum_j \hat{a}_{ij}$, derives from a potential. However, it is not guaranteed that real data will satisfy this condition. In general, we would expect vector fields to have rotational contributions. According to the developed theory, this would imply that the species dynamics are not compatible with an equation of the form given by Eq.~\eqref{eq:def_model}; or that the errors introduced by finite $\Delta x$, $\Delta t$, and $t$ are too big to apply this method. 

\section{Self-interactions in the geometric Brownian model}\label{AP_sec:int_GBM}
In this section, we compute self-interactions over paths generated with the geometric Brownian model,
\begin{equation}
    dX_t = v\,X_t\,dt + D\,X_t\,dW_t,
\end{equation}
with solution
\begin{equation}\label{AP_eq:GBM_solution}
    X_t = X_0\, e^{(v-\frac{D^2}{2})t+D\,W_t}.
\end{equation}
In this case, the instantaneous per-capita growth rate measured from a realization of the process with $X_0=x_0$, 
\begin{equation}
    g^{(E)}(t,\Delta t|x^{(0)}) = \frac{X(t+\Delta t|x^{(0)})-X(t|x^{(0)})}{\Delta t\,X(t|x^{(0)})},
\end{equation}
can be computed exactly using \eqref{AP_eq:GBM_solution},
\begin{equation}\label{AP_eq:GBM_growth_rate}
    g^{(E)}(t,\Delta t|x^{(0)}) = \frac{1}{\Delta t}\left[e^{(v-\frac{D^2}{2})\Delta t+D\,\Delta W_t}-1\right] ,
\end{equation}
where 
\begin{equation}
    \Delta W_t = W_{t+\Delta t}-W_t.
\end{equation}
The empirical measure of interactions can also be evaluated from eq.\eqref{AP_eq:GBM_growth_rate} as
\begin{align}\label{AP_eq:interactions_GBM_model}
    &\mathcal{M}(t,\Delta x,\Delta t) = \nonumber
    \\& g^{(E)}(t,\Delta t|x^{(0)}+\Delta x)-g^{(E)}(t,\Delta t|x^{(0)}) =\nonumber
    \\ & \frac{1}{\Delta t}\left[e^{(v-\frac{D^2}{2})\Delta t+D\,\Delta W_t}-1\right]-\frac{1}{\Delta t}\left[e^{(v-\frac{D^2}{2})\Delta t+D\,\Delta \tilde{W}_t}-1\right],
\end{align}
where $W_t$ and $\tilde{W}_t$ represent two distinct realizations of the Wiener process. For small $\Delta t$, Eq.~\eqref{AP_eq:interactions_GBM_model} can be further simplified to 
\begin{align}\label{AP_eq:interactions_GBM_model_approx}
    &\lim_{\Delta t\to 0}\mathcal{M}(t,\Delta x,\Delta t) = \lim_{\Delta t\to 0}\frac{D}{\Delta t}\left[\Delta W_t-\Delta \tilde{W}_t\right]=\lim_{\Delta t\to0} \sqrt{\frac{2}{\Delta t}}D\,\mathcal{N}(0,1).
\end{align}

\section{Effect of noise correlations in growth rates estimates}\label{AP:int_correlated}
In this section, we study per-capita growth rate estimates measured over trajectories generated by a process with colored noise: \begin{equation}\label{AP_eq:colored_process} dX_t = \left(v+O_t\right)\,X_t\,dt, \end{equation} where $O_t$ is an Ornstein-Uhlenbeck process, \begin{equation} O_t =O_0 , e^{-\frac{t}{\tau}}+D\int_0^t e^{-\frac{t-s}{\tau}}dW_s. \end{equation} The process in Eq.\eqref{AP_eq:colored_process} has the solution \begin{equation} X_t = X_0\, e^{v,t+\int_0^tO_s,ds}. \end{equation} Inserting this solution into Eq.\eqref{eq:def_growth_rate_E}, we obtain the empirical growth rate over stochastic paths: 
\begin{equation} 
g^{(E)}i\left[\Delta t,t,x_0\right] = \frac{e^{v\,\Delta t+\int_t^{t+\Delta t}\,O_s ,ds}-1}{\Delta t}. \end{equation}
The limit $\Delta t\to0$ can be safely computed, 
\begin{equation}
\lim_{\Delta t\to0}g^{(E)}_i\left[\Delta t,t,x_0\right] = v+ O_t, 
\end{equation}
which is a well-defined random quantity corresponding to the deterministic growth rate with Gaussian perturbation.

If instead we generate data using uncorrelated noise, \begin{equation}\label{AP_eq:uncorrelated_process} 
dX_t = \left(v \,dt+dW_t\right)\,X_t, \end{equation} 
we obtain 
\begin{equation} 
g^{(E)}_i\left[\Delta t,t,x_0\right] = \frac{e^{v\,\Delta t+W_{t+\Delta t}-W_t}-1}{\Delta t}, 
\end{equation} 
which has no well-defined limit as $\Delta t\to0$, since \begin{equation}
\lim_{\Delta t\to 0}g^{(E)}_i\left[\Delta t,t,x_0\right] = v+\lim_{\Delta t\to 0}\frac{\Delta W_t}{\Delta t}\notin \mathbb{R}. \end{equation}

\end{document}